%% file: main.tex
\documentclass[nofootinbib,superscriptaddress,a4paper,aps,prx,twocolumn,longbibliography, nobalancelastpage]{revtex4-2}

\usepackage{graphicx}
\usepackage{dcolumn}
\usepackage{bm}
\usepackage{amsfonts}
\usepackage{amsmath}
\usepackage{amssymb}
\usepackage{subfigure}
\usepackage{tikz}
\usetikzlibrary{decorations.pathreplacing,decorations.pathmorphing}
\usepackage{booktabs}

\usepackage[export]{adjustbox}
\newsavebox{\imagebox}

\usepackage{hyperref}
\usepackage{xcolor}
\hypersetup{
    colorlinks,
    linkcolor={red!50!black},
    citecolor={blue!50!black},
    urlcolor={blue!80!black}
}

\newcommand{\lr}[1]{\ensuremath{\left( #1 \right)}}
\newcommand{\cop}[1]{\ensuremath{\hat{{#1}}^\dagger}}
\newcommand{\aop}[1]{\ensuremath{\hat{{#1}}}}

\newcommand{\ket}[1]{\ensuremath{\lvert #1 \rangle}}
\newcommand{\bra}[1]{\ensuremath{\langle #1 \rvert}}
\newcommand{\ketbra}[2]{\ensuremath{\lvert #1 \rangle\langle #2 \rvert}}

\newcommand{\expval}[2]{\ensuremath{\langle #1 \rangle_{#2} }}

\begin{document}

\title{Quantum computer-aided design of quantum optics hardware}
\author{Jakob~S.~Kottmann}
\email{jakob.kottmann@utoronto.ca}
\affiliation{Chemical Physics Theory Group, Department of Chemistry, University of Toronto, Canada.}
\affiliation{Department of Computer Science, University of Toronto, Canada.}

\author{Mario Krenn}
\email{mario.krenn@univie.ac.at}
\affiliation{Chemical Physics Theory Group, Department of Chemistry, University of Toronto, Canada.}
\affiliation{Department of Computer Science, University of Toronto, Canada.}
\affiliation{Vector Institute for Artificial Intelligence, Toronto, Canada.}

\author{Thi Ha Kyaw}
\affiliation{Chemical Physics Theory Group, Department of Chemistry, University of Toronto, Canada.}
\affiliation{Department of Computer Science, University of Toronto, Canada.}

\author{Sumner Alperin-Lea}
\affiliation{Chemical Physics Theory Group, Department of Chemistry, University of Toronto, Canada.}
\affiliation{Department of Computer Science, University of Toronto, Canada.}

\author{Alán Aspuru-Guzik}
\email{aspuru@utoronto.ca}
\affiliation{Chemical Physics Theory Group, Department of Chemistry, University of Toronto, Canada.}
\affiliation{Department of Computer Science, University of Toronto, Canada.}
\affiliation{Vector Institute for Artificial Intelligence, Toronto, Canada.}
\affiliation{Canadian  Institute  for  Advanced  Research  (CIFAR)  Lebovic  Fellow,  Toronto,  Canada}
\date{\today}

% \author{Jakob S. Kottmann}
% \address{Department of Chemistry, University of Toronto}
% \address{Department of Computer Science, University of Toronto}

% \author{Mario Krenn}
% \address{Department of Chemistry, University of Toronto}
% \address{Department of Computer Science, University of Toronto}

% \author{Thi Ha Kyaw}
% \address{Department of Chemistry, University of Toronto, Toronto}
% \address{Department of Computer Science, University of Toronto}

% \author{Sumner Alperin-Lea}
% \address{Department of Chemistry, University of Toronto}

% \author{Al\'{a}n Aspuru-Guzik}
% \address{Department of Chemistry, University of Toronto, Toronto, Ontario M5G 1Z8, Canada}
% \address{Department of Computer Science, University of Toronto}
% \address{Vector Institute for Artificial Intelligence, Toronto}
% \address{Canadian Institute for Advanced Research, Toronto}
% \ead{alan@aspuru.com}

\begin{abstract}
 The parameters of a quantum system grow exponentially with the number of involved quantum particles. Hence, the associated memory requirement to store or manipulate the underlying wavefunction goes well beyond the limit of the best classical computers for quantum systems composed of a few dozen particles, leading to serious challenges in their numerical simulation. This implies that the verification and design of new quantum devices and experiments are fundamentally limited to small system size. It is not clear how the full potential of large quantum systems can be exploited. Here, we present the concept of quantum computer designed quantum hardware and apply it to the field of quantum optics. Specifically, we map complex experimental hardware for high-dimensional, many-body entangled photons into a gate-based quantum circuit. We show explicitly how digital quantum simulation of Boson sampling experiments can be realized. We then illustrate how to design quantum-optical setups for complex entangled photonic systems, such as high-dimensional Greenberger-Horne-Zeilinger states and their derivatives. Since photonic hardware is already on the edge of quantum supremacy and the development of gate-based quantum computers is rapidly advancing, our approach promises to be a useful tool for the future of quantum device design.
\end{abstract}

\maketitle 
%%%%%%%%%%%%%%%%%%%%%%%%%%%%%%%%%%%%%%%%%%%%%%%%%%%%%%%%%%%%%%%%%%%%%%%%%%%%%%%%%%%%%%%%%%%%%%%%%%%%%%%%%%%%%%

\section{Introduction}
Photonic systems are highly flexible and controllable for small to medium-sized quantum systems, and offer resilience against decoherence~\cite{pan2012multiphoton,flamini2018photonic}. These properties make them a first choice in many proof-of-concepts in quantum information science. Examples include observations of fundamental quantum properties, such as indefinite causal orders~\cite{rubino2017experimental}, early demonstrations of Wigner's friend paradox~\cite{proietti2019experimental,bong2019testing}, high-dimensional quantum communication systems such as quantum key distribution~\cite{mirhosseini2015high,sit2017high}, entanglement swapping~\cite{zhang2017simultaneous}, quantum teleportation~\cite{luo2019quantum} and experimental quantum machine learning~\cite{pepper2019experimental, rocchetto2019experimental} and new propositions for quantum technologies~\cite{kristensen2019artificial, cao2017quantum, anschuetz2019realizing, Alcazar_2020, Benedetti_2018, lamata2020}.\\

While quantum experiments historically have been designed by experienced human experts, their non-intuitive nature has led to the emergence of computational methods for designing quantum experiments~\cite{krenn2016automated, knott2016search, arrazola2018machine, wallnofer2019machine, nichols2019designing, zhan2020experimental, gubarev2020improved, krenn2020computer}. However, as the dimension of state space grows exponentially with the number of photons, this approach is limited to small systems. Consequently, while the abilities of photonic hardware constantly improve~\cite{zhong201812,wang201818,llewellyn2019chip,lu2020three,wang2020integrated}, there is no efficient computational method that can take advantage of the vast resources provided by these systems. Furthermore, photonic quantum supremacy experiments are close to the point where they cannot be calculated with classical hardware~\cite{wang2019boson}. How can one verify their correct execution when such calculations step beyond the point of classical calculation?\\

Here we illustrate a solution to solve the verification and the design processes of quantum optical setups. We demonstrate how quantum optical systems can be recast in the language of digital quantum computers and use the state-of-the-art simulators of quantum computers to design experiments for complex multi-photon entangled quantum systems. Furthermore, we showcase the {quantum simulation} for one of the first demonstrations of Boson sampling~\cite{crespi2014}, illustrating that digital quantum computers can function as witnesses for photonic quantum supremacy experiments that are expected in the near future.\\

Due to the rapid progress in the development of gate based quantum computers in the recent years~\cite{erhard2019characterizing, arute2019quantum}, we estimate that the design of photonic hardware with quantum computers will become a realistic scenario in the near future.
In the meantime, the optimization strategies presented here could also serve as valuable benchmarks complementary to quantum chemistry~\cite{mccaskey2019quantum}.
While the latter mainly focuses on determining an energy for an unknown ground state, the optimization of an optical setup focuses on determining its parameters for a desired target quantum state.\\

We propose the design and {simulation} of general quantum hardware as a new application for quantum computers. In this manuscript, we focus on photonic quantum hardware by translating optical elements and measurement techniques into gate based quantum computers language. In a separate paper, some of us target the design of efficient superconducting qubit architectures by translating the corresponding Hamiltonians into a digital quantum circuit~\cite{kyaw2020}.

%%%%%%%%%%%%%%%%%%%%%%%%%%%%%%%%%%%%%%%%%%%%%%%%%%%%%%%%%%%%%%%%%%%%%%%%%%%%%%%%%%%%%%%%%%%%%%%%%%%%%%%%%%%%%%
\section{Quantum Simulation of Optical Elements}\label{sec:circuits}

\begin{figure*}[ht]
    \centering
    \includegraphics[width=0.80\textwidth]{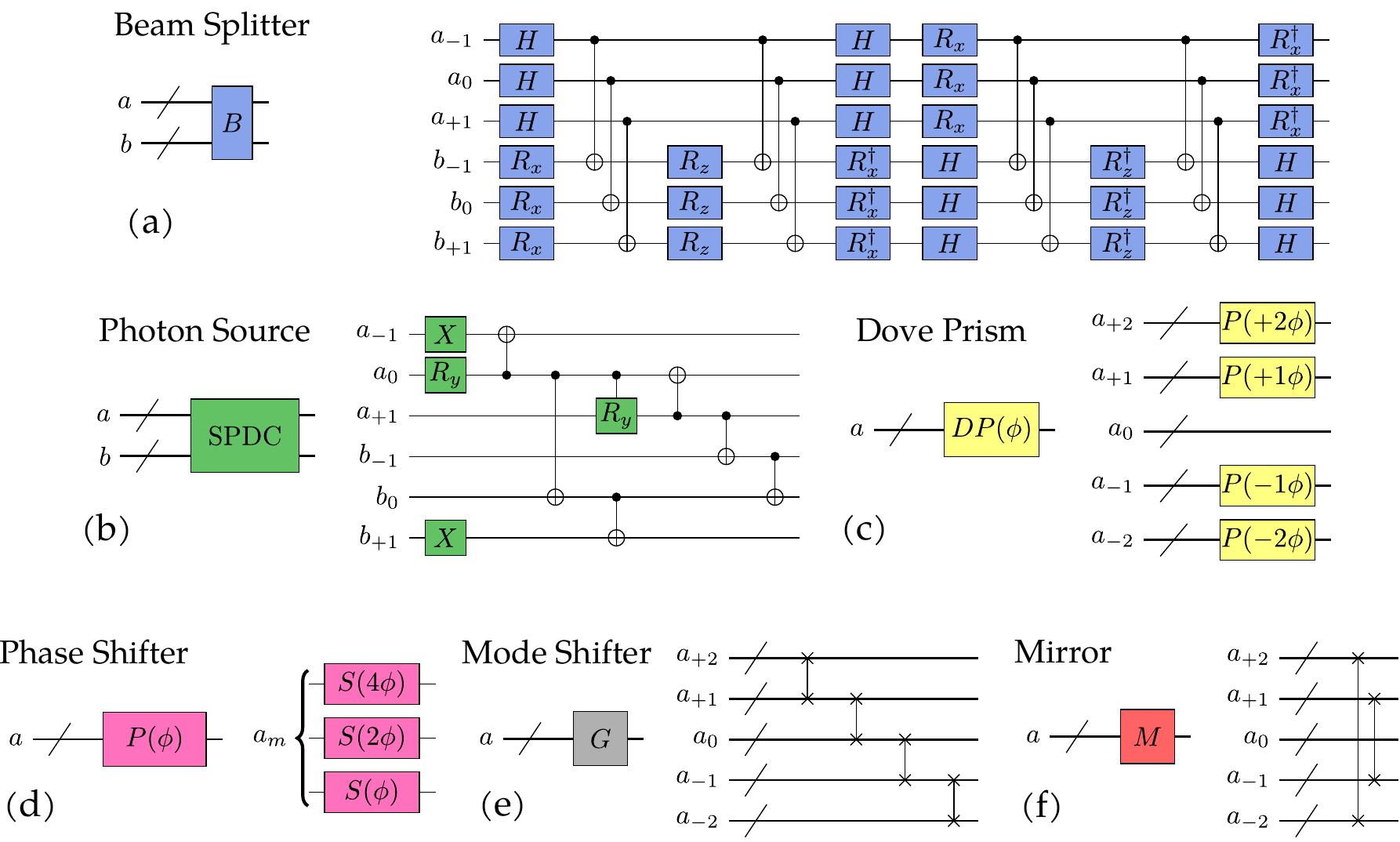}
    \caption{\textbf{Quantum circuits for multi-photonic high-dimensional quantum optics.} Optical paths are denoted by $a,b,c$ while internal mode numbers are denoted by subscripts. Here we use the orbital angular momentum of photons as a high dimensional degree of freedom. In general, this approach can be applied to any discrete high dimensional quantum numbers. Each internal mode is represented by several qubits representing the photon occupation number (see Supplementary Information for additional details). (a) Example of a beam splitter used in Fig.~\ref{fig:332_example} where each internal mode is represented by one qubit. The general multi-photon beam splitter (see Fig.~\ref{fig:crespi_example}) is constructed with a Trotter expansion and is too large to show here. (b) Direct emulation of a high-dimensional entangled photon state created by spontaneous parametric down conversion in a nonlinear crystal. (c) Mode dependent phase shifter (Dove prism) implemented as multiple phase shifters acting on the corresponding modes. (d) Mode independent phase shifter where the photonic occupation number is encoded in 3-qubit binary representation (up to 7 photons per mode). (e) Cyclic approximation to a mode shifter (hologram) implemented by swap gates (each  swap denotes a swap operation on all the qubits which represent the mode). (f) Mirror implemented by swap gates.}\label{fig:overview}
\end{figure*}

In the following, we will explain the mapping from quantum optics onto quantum circuits.
A quantum optical setup consists of multiple optical path modes (paths) which can be occupied by multiple photons with additional internal degrees of freedom (modes) like for example orbital angular momentum of light~\cite{rubinsztein2016roadmap,padgett2017orbital,erhard2018twisted}.
The photonic occupation number of each internal degree of freedom is represented by a set of qubits.
We will use binary encoding and we refer to the Ref.~\cite{nicolas2019} for detailed analyses of other encodings (see also Refs.~\cite{sabin2020digital, somma2003} for unary encodings).
In this representation, the number of qubits needed to represent an optical setup is given by
\begin{eqnarray*}
{N_\text{qubit} = N_\text{modes}\times N_\text{paths}\times \lceil\log_2\lr{ N_\gamma }\rceil},
\end{eqnarray*}
where $N_\gamma$ is the maximum number of photons in one mode and we have used the integer ceiling function.
With this encoding, a basis state of the photonic setup can be represented as $\bigotimes_p \bigotimes_m \ket{n_{m,p}}$  
where $p,m,n$ represent path, mode and number of photons, respectively.
Take as an example a setup with $N_\text{paths}=1$ path, and $N_\text{modes}=3$ internal degrees of freedom, denoted as $\left\{-1, 0, 1\right\}$, where each can be occupied by up to $N_\gamma=3$ photons.
A state in this setup can then be represented by $N_\text{qubit}=6$ qubits. Assuming that $2$ of the photons occupy the mode $-1$ and $1$ photon occupies mode $1$, the state can then be denoted as
\begin{eqnarray*}
   \ket{2_{-1,a}, 0_{0,a}, 1_{+1,a}} \xrightarrow[]{\text{qubits}} \ket{10}_{-1,a} \otimes \ket{00}_{0,a} \otimes \ket{01}_{+1,a}.
\end{eqnarray*}
These photonic states can be transformed by optical elements which can be represented by digital quantum gates.
In Fig.~\ref{fig:overview} we show gate based representations of important optical elements for high-dimensional quantum optics and provide further details in the appendix.

\subsection{Implementation}
One of the advantages of simulating the optical setups on a digital quantum computer is the direct access to gradients of parametrized elements within a fully automatically differentiable framework.~\cite{schuld2019, bergholm2018pennylane}
This fulfills all necessary conditions to replace the classical simulation module of topological optimizers such as \texttt{theseus}~\cite{krenn2020conceptual} as illustrated in Fig.~\ref{fig:big_picture}.
We will illustrate the underlying framework here using the phase shifter of Fig.~\ref{fig:overview} as an explicit example. The phase shifter is shifting the relative phase of the photonic state depending on the number of photons occupying the photonic path onto which the element acts. Using the Bosonic number operator, a phase shifter acting on path mode $a$ can then be written as
\begin{eqnarray}
    PS\left(\phi\right) = e^{i a^\dagger a}
\end{eqnarray}
with the associated relative phase $e^{i\phi}$.
With the mapping of Fig.~\ref{fig:mapping} an $n$-qubit implementation of the phase shifter can be realized by a collection of single qubit operations
\begin{eqnarray}
    P\lr{\phi} \xrightarrow{n-qubits} S\lr{2^{n-1}\phi} \otimes S\lr{2^{n-2}\phi} \otimes \dots S\lr{\phi},
\end{eqnarray}
where the $S(\phi) = e^{-i\frac{\phi}{2} \left( \sigma_z - 1 \right)}$ gate adds the phase $e^{i\phi}$ to the state $\ket{1}$ and leaves $\ket{0}$ invariant (see the appendix for more details). In the optimization of a quantum optical setup, we aim to optimize the fidelity of the state, created by the setup, with a specific target state. In the next section (Eqs.~\eqref{eq:fidelity_simple} and ~\eqref{eq:fidelity_ps}) we will construct this fidelity as a function of expectation values $E= \expval{H}{U(\phi)}$ depending on the phase shifter parameter $\phi$ through the unitary $U(\phi)$ that encodes the quantum optical setup, and a Hamiltonian $H$ that encodes the measurements. In order to compute the gradient of the fidelity with respect to the phase shifter parameter $\phi$, we need the gradients of the expectation value $\partial_\phi E$. Following Schuld \textit{et.al.}~\cite{schuld2019} those gradients can be obtained with the parameter-shift rule, which allows the evaluation of the \textit{analytical} gradients via a finite-difference like procedure as $ \partial_\phi E = E(\phi-\frac{\pi}{4r}) + E(\phi+\frac{\pi}{4r}) $. In order for this technique to be applicable, the generator of the parametrized quantum gate is required to have only two distinct eigenvalues with distance $2r$. Quantum gates that do not fulfill this condition can be decomposed into more primitive gates, which allows to evaluate their gradients by combining the parameter-shift rule with the product rule of calculus (see~\cite{tequila} for illustrations). 
In our explicit example the phase shifter is represented by a set of $S(\phi)$ gates with the generator $\frac{1}{2}\left( \sigma_z - 1 \right)$. These gates fulfills the condition ($r=\frac{1}{2})$ and no further decomposition is necessary. Gradients of the other parametrized optical elements, such as the beam splitter or the heralding process are obtained in an analogue fashion. The parametrized part of the beam splitter is represented by single qubit rotations which fulfill the requirements for the parameter-shift rule and also do not require further decomposition. In our example in Fig.~\ref{fig:332_example} the heralding process contains a controlled rotation that needs to be further decomposed (see~\cite{tequila} for an explicit example). In our implementation we use \textsc{tequila}~\cite{tequila}, a high level python package that  automatizes the illustrated gradient compilation and associated gate decompositions. We refer to Refs.~\cite{tequila, kottmann2020feasible} for more details on the implementation of the automatically differentiable framework.
Note that the evaluation of the gradient in this way would not be possible on an actual quantum optical setup, since the decomposition of the optical elements into more primitive gates is not possible as those elements are already the primitive building blocks. This constitutes one of the main advantages of our approach.

%%%%%%%%%%%%%%%%%%%%%%%%%%%%%%%%%%%%%%%%%%%%%%%%%%%%%%%%%%%%%%%%%%%%%%%%%%%%%%%%%%%%%%%%%%%%%%%%%%%%%%%%%%%%%%
\section{Quantum Optical Setups}

\begin{figure}
    \centering
    \includegraphics[width=0.4\textwidth]{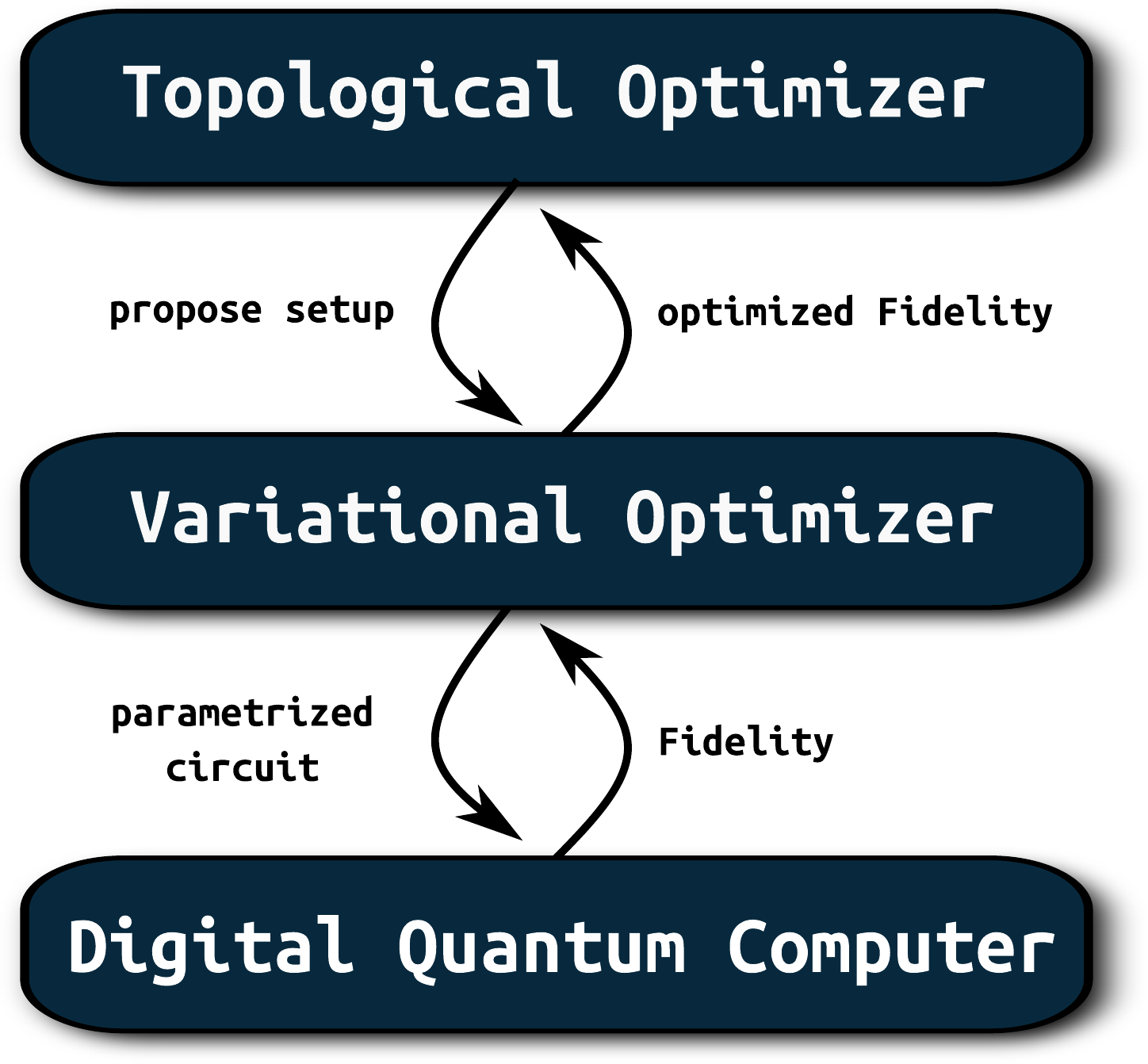}
    \caption{\textbf{Quantum computer-aided design of quantum optics hardware}: Proposed usage of the developed techniques in combination with topological optimizers like \textsc{melvin}~\cite{krenn2016automated} or \textsc{theseus}~\cite{krenn2020conceptual}. Within those optimizers, the (classical) computation of the involved fidelities is currently the computational bottleneck. In this work we propose to replace this classical subroutine with a quantum algorithm that can evaluate and optimize the corresponding fidelities.}
    \label{fig:big_picture}
\end{figure}

In the following we will describe how to optimize the fidelity of a parametrized optical setup with a specific target state on a digital quantum computer. The goal is to determine the optimal parameters of a quantum optical setup for quantum communication, quantum metrology and experiments testing foundations of quantum physics. Optimizing the quantum optical setup on a universal gate based quantum computer has unique advantages, two of which are particularly relevant here: First, the initial state preparation can be deterministic, in contrast to widely used probabilistic photon state sources. Second, the access to universal gates allows the evaluation of analytical gradients as illustrated in the last section as well as more efficient measurement protocols. 
Our concrete example results in measuring only the occurrence of one specific product state as a proxy for the fidelity of a complex entangled state.
In the future one could imagine large parametrized optical setups simulated on quantum computers.
Ideally, the simulation would reduce the setup size by optimizing parameters of specific elements such that they are equivalent to the application of the identity, allowing the removal of said elements. 
In this case the optimized topology would emerge. Recently some of us developed classical graph based optimization methods for quantum optics. Here, the above proposed reduction of the large setups through parameter optimization could already be realized with a classical simulation.\cite{krenn2020conceptual} We believe that the full potential of the techniques developed in this work will be reached in combination with those topological optimization methods. In Fig.~\ref{fig:big_picture} we illustrate possible combinations of classical topological optimizers with the techniques proposed here and add further details in the appendix.
The quantum part can, for example, be an efficient sub-module of the overall topological optimization.
{Note, that in the following proof of principle illustration, we chose a setup from the literature that incorporates heralding and post-selection. Therefore choosing a setup that is not trivial, but also small enough to envision execution on currently emerging ion-trap or superconducting quantum hardware. Recent improvements on the topological optimizers, developed simultaneously with this work, directly led to solutions of previously unanswered questions in experimental quantum optics.~\cite{krenn2020conceptual} Those improvements where however mainly concerned with the topological part of the optimization and the simulation of the optical setup is currently becoming the bottleneck. At this point, a powerful digital quantum computer can step in, and take over the simulation of the optical setups within the topological optimization. Compared to other scientific fields like quantum chemistry, the necessary classical simulation of quantum optical setups is not as deeply explored, so that further improvements in this direction bear a high potential. Although we are primarily concerned with quantum simulation in this work, our implementation within \textsc{tequila} provides an ideal testbed for improved classical optimizers and simulators. One example might be state of the art tensor-network based contraction methods~\cite{gray2018quimb} and we are currently exploring this possibilities.}\\

The quantum part of the optimization is performed in the spirit of variational quantum eigensolvers (VQE's) originally proposed to variationally approximate eigenstates of a given Hamiltonian~\cite{peruzzo2014}.
In this work, we use a variational algorithm to optimize fidelities for a given target state which can be written as the expectation value
\begin{eqnarray}
    F_\Psi  =  \lvert \bra{\Psi}\ket{\Phi} \rvert^2 = \bra{\Phi}H\ket{\Phi},\label{eq:fidelity_simple}
\end{eqnarray}
with the Hamiltonian
$
    H = \ket{\Psi}\bra{\Psi}\label{eq:|Psi><Psi|},
$
and where $\Psi$ is the desired target state.
Depending on $\Psi$, the number of measurable components (tensor products of Pauli matrices) in the Hamiltonian, can grow large.
One proposed way to reduce the number of measurements is to group the Hamiltonian into commuting cliques~\cite{izmaylov2020, yen2019measuring, verteletskyi2019measurement}, a technique which could be applied here in the same way.
Since in contrast to most VQE optimizations the target state is known here, we can measure the Hamiltonian directly by using the unitary $U_\Psi$ that prepares the target state and measure the transformed projector
\begin{eqnarray}
    P_0 = U_\Psi^\dagger H U_\Psi = \ket{00\dots0}\bra{00\dots0} = \bigotimes_j \frac{1}{2}\lr{1 + Z_j}\label{eq:def_Htilde},
\end{eqnarray}
where $Z_j$ denotes the Pauli $Z$-matrix on qubit $j$.
This technique allows us to further optimize for target states where only the unitary operation on the digital quantum machine is known, this includes states which are themselves optimized by variational quantum algorithms beforehand.
The expectation value of this Hamiltonian can then be estimated by measuring all qubits in the computational basis and counting the ``all-zero'' results where the probability of measuring them is directly proportional to the desired fidelity.

\subsection*{Optimization of a heralded, post-selected state}
Two common measurement based preparation strategies in quantum optics are heralding and post-selection.
Heralding means that the measurement of a trigger photon in an ancillary path determines the success of the preparation.
This projection of the state by measuring the ancillary path can be represented in the same way as above by measuring the following projector
\begin{eqnarray*}
    P_p = U_pP_0U_p^\dagger,
\end{eqnarray*}
where the $U_p^\dagger$ transforms the state $\ket{p}$ in which the trigger photon is measured into the $\ket{0\dots0}$ state.
Note that the state in which the trigger photon is measured, can be optimized when the unitary $U_p$ is parametrized (see Fig.~\ref{fig:332_example} for an example).
In addition, the generated states can be restricted by post selection that only outcomes of the experiment are counted with one photon in each measured path.
Since the post-selection projector acts on the same paths as the Hamiltonian representing the fidelity, it is not possible to directly use the transformed Hamiltonian of \eqref{eq:def_Htilde} to reduce the number of measurements.
If one can afford using twice the number of qubits the approach of Ref.~\cite{mazzola2019}, can be applied, where the information about the photon occupation number is transferred to additional ancillary qubits by a series of controlled not operations.
Depending on the specific setup, the number of ancillary qubits can be reduced by constructing an efficient encoding $E$ (see Fig.~\ref{fig:332_example} for an example).
The post-processing is then carried out over the additional registers.
For both methods, it is important to normalize the fidelity in order to ensure that the parametrization leading to the highest fidelity \textit{after} applying both projectors is the global minimum of the loss function.
The fidelity one needs to optimize is
\begin{eqnarray}
\mathcal{F}_\Psi = \frac{ \bra{\Phi} P_p \otimes HP_1\ket{\Phi}}{\bra{\Phi} P_p \otimes P_1  \ket{\Phi}}, \label{eq:fidelity_ps}
\end{eqnarray}
where $P_1$ projects onto the one-photon subspace.

\subsection*{Example: Optimization of a post-selected heralded \textit{332-state}}\label{sec:332_example}

\begin{figure*}[ht]
    \centering
    \includegraphics[width=0.99\textwidth]{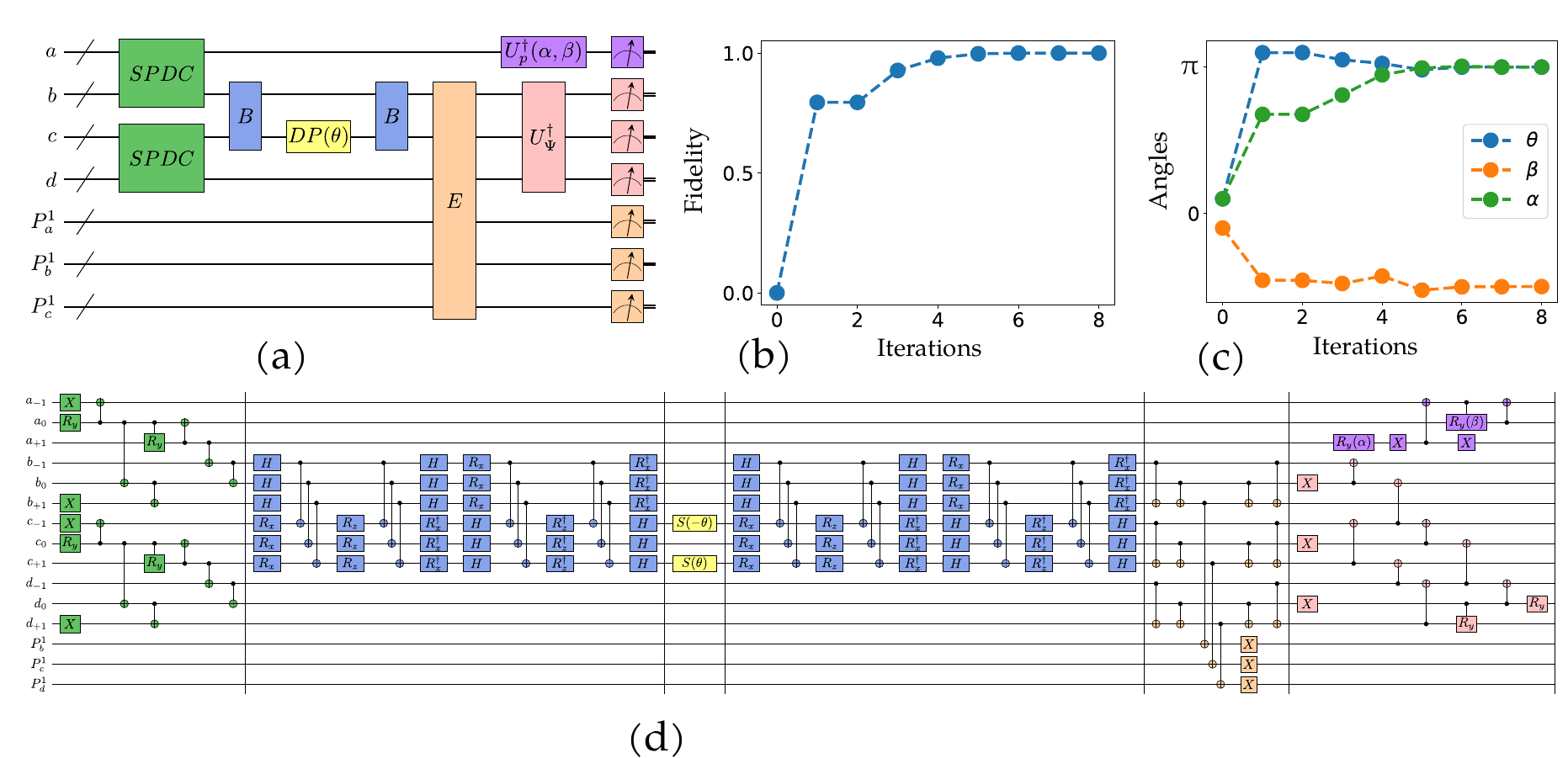}
    \caption{\textbf{Optimization of an asymmetric high-dimensional entangled state} (a) Abstract representation of the optimization setup. Optical elements are shown in green, blue and yellow and are described in Fig.~\ref{fig:overview}. $U_p^\dagger(\alpha, \beta)$ transforms a parametrized photonic qutrit to the computational zero-state which acts as a trigger for the three-photon state in paths $b,c,d$. The encoding $E$ emulates the post-selection by transferring information about the photon number in paths $b,c,d$ to auxiliary qubits $P^1_b,P^1_c,P^1_d$ leading them to be in the zero-state for valid configurations. Similar as the trigger in $a$, $U^\dagger_\Psi$ transforms the target state of \eqref{eq:332-state} into the zero-state. The probability of measuring the overall zero-state directly corresponds to the heralded and post-selected fidelity of \eqref{eq:fidelity_ps} with the target state $\Psi$. The setup is parametrized with three angles (one for the Dove prism and two for the trigger). (b-c) Optimization of the setup with the 'BFGS' optimizer. (d) Explicit circuit for the setup in (a).       }
    \label{fig:332_example}
\end{figure*}

As an instructive example we will show here the optimization of a parametrized quantum optical setup targeted to produce a so called \textit{332-state}~\cite{huber2013structure,huber2013entropy}, a state with high-dimensional multipartite entanglement, where the first two photons are entangled over all three paths and the third over two paths. Using (-1, 0, +1) as internal degrees of freedom a \textit{332-state} can be written as
\begin{eqnarray}
    \ket{\text{\textit{332}}} =& \frac{1}{\sqrt{3}} \left( \ket{1_{-1,a},1_{1,b},1_{-1,c}} \right. \\ 
    &+ \ket{1_{0,a}, 1_{0,b}, 1_{0,c}} + \ket{1_{-1,a}, 1_{-1,b}, 1_{1,c}} \left.\right).\nonumber \label{eq:332-state}
\end{eqnarray}
The preparation of the \textit{332-state} is a subset of the setup to prepare a multidimensional GHZ state and the setup with the right parametrization has been found by using automatic generation and search algorithms~\cite{krenn2016automated} where a variational quantum optimization shown in Fig.~\ref{fig:big_picture} was not applied. After it was found by computer-aided design, the proposed state preparation setup could be demonstrated experimentally~\cite{malik2016multi, erhard2018experimental}
The parametrized setup which can generate a post-selected heralded \textit{332-state} is shown in Fig.~\ref{fig:332_example} where the state is created in the photonic paths $b,c,d$ and a measurement of the photon in path $a$ is used as a trigger for the successful preparation of the state in paths $b,c,d$.
While the preparation of the state is non-trivial with a quantum optical setup, it can be directly prepared on a digital quantum computer as shown in  Fig.~\ref{fig:332_example}.
Another advantage of the digital quantum simulation is the direct generation of initial states (here high-dimensional Bell states), which, on real photonic devices, have to be created in a probabilistic fashion with spontaneous parametric down-conversion (SPDC).
Successful pair generation events through SPDC are rare, and the count rates are usually measured in counts per hour~\cite{krenn2020computer}. This is one of the reasons that make direct optimization of the photonic hardware a less desirable task.
In other words, on the digital quantum computer, we simulate only the runs of the quantum optics experiment with successful state initialization, circumventing the associated waiting times that come with the low count rates for the original photonic device.
For this example, it is possible to approximate each internal degree of freedom by a single qubit and use an efficient encoding $E$ for the implementation of the single-photon projector leading to an overall circuit involving $15$ qubits.
The encoding $E$ flips the ancillary qubits ($P^1_b, P^1_c, P^1_d$)  assigned to each path ($b,c,d$) if the path has one or three photons and the post-selection process is encoded by counting only measurements where all three ancillary qubits are zero.
Notice that the three photon case does not influence the result in this example since having three photons in all three paths is not possible in this setup.
Note as well that due to the encoding of the post-selection process, the fidelity with the target state as well as the measurement process of the trigger photon in path $a$ is again the same as in \eqref{eq:def_Htilde}. That means measuring all $15$ qubits in the computational basis and counting the overall all-zero results directly leads to the desired heralded and post-selected Fidelity of \eqref{eq:fidelity_ps}.
For the initialization, we choose angles close to zero with varying signs. We show one particular optimization in Fig.~\ref{fig:332_example}(c).

\subsection*{Example: Boson-Sampling}

\begin{figure*}[ht]
    \centering
    \includegraphics[width=0.80\textwidth]{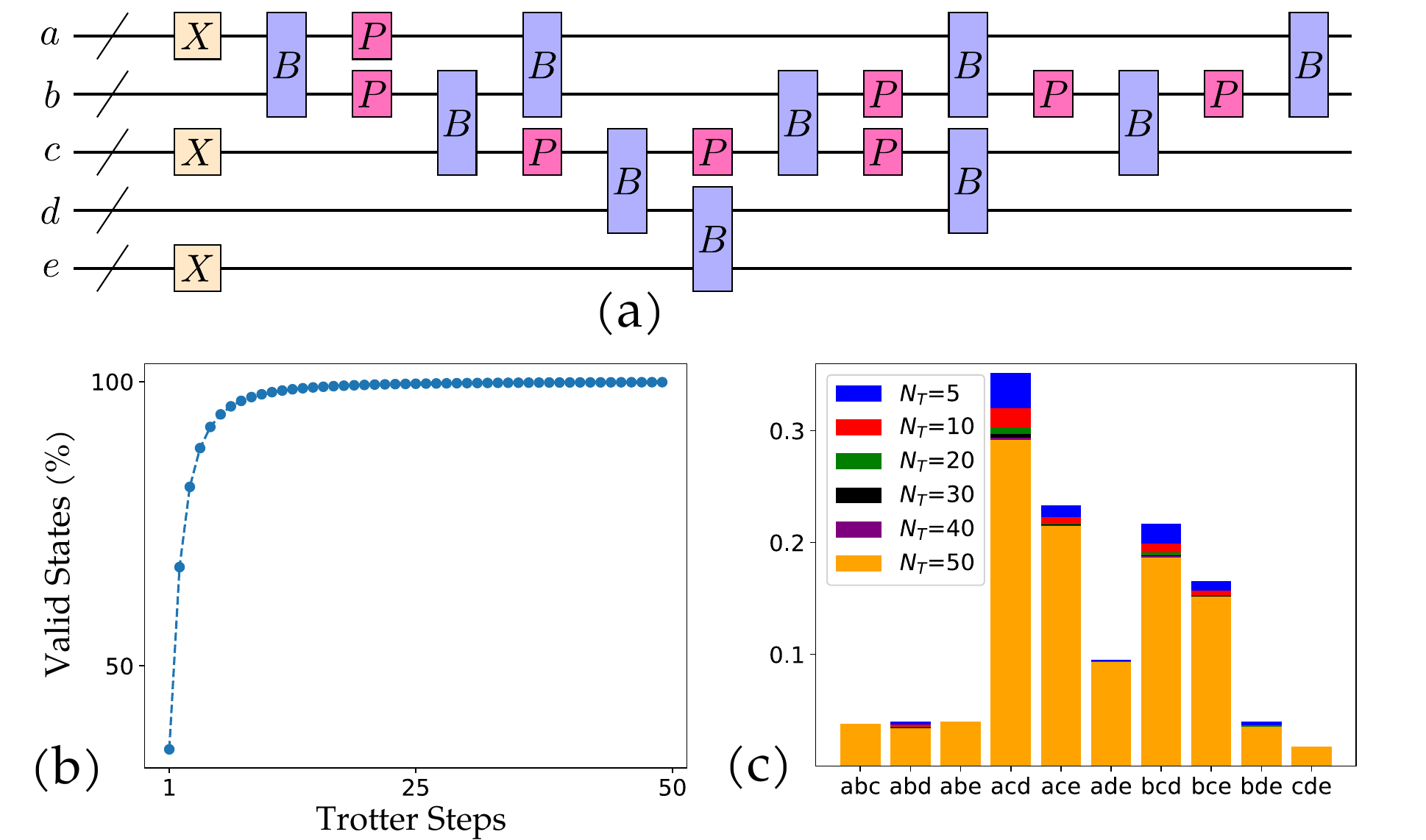}
    \caption{\textbf{Digital quantum simulation of a Boson sampling experiment}. (a) In the abstract representation of the setup each path is represented by two qubits (allowing to represent 0-3 photons in each path). The setup consists of beam-splitters (B) and phase shifters (P) and is initialized with three photons in paths $a$, $c$ and $e$ ($\ket{1_a0_b1_c0_d1_e}$); The bit-flip $X$ gates only operate on one of the two qubits. (b) Percentage of physically valid states -- meaning configurations of the simulated wave function obeying photon number conservation and only having three photons in total -- as an indicator of the error introduced by the Trotter expansion. (c) Simulated distribution of three photon states with each photon in a separate path. At 10 Trotter steps the error with respect to the exact quantum optical setup is about 2 percent, and consistent with the results presented in~\cite{crespi2014}. 
    }
    \label{fig:crespi_example}
\end{figure*}

A potentially interesting application for simulating quantum optical setups on a digital quantum computer is the quantum simulation of Boson sampling experiments, which are currently approaching a size inaccessible for classical computers.
Boson sampling~\cite{aaronson2011computational} is one major candidate for demonstrating quantum supremacy on near-term quantum devices.
There is no doubt that it will become challenging to simulate Boson sampling devices as the system size grows.
Several classical validation methods exist which, though unable to confirm Boson sampling directly, may be used to rule out sampling from more classically accessible distributions.
Examples involve row norm estimates~\cite{aaronson2013, Spagnolo2014, clifford}, Bayesian analysis~\cite{wang2019boson}, Kolmogorov–Smirnov tests~\cite{neville2017classical} and machine learning approaches~\cite{agresti2019pattern}.
The approach here however simulates the underlying optical sampling setup directly providing the possibility to manipulate the quantum state through unitary operations at different stages of the process.
At and beyond the quantum supremacy threshold, simulation with a classical computer will, for the general case, become impossible in practice.
A powerful digital quantum computer could at this point step in and reproduce the process.
Apart from solely showing quantum supremacy, there are proposals to employ them for more valuable computations like the generation of Frank-Condon spectra of molecules~\cite{huh2015}, for which cross simulation will be a useful tool. It is currently unclear if specialized quantum algorithms on digital machines could achieve the same task more efficiently. Here our mapping would provide a first test for prospective candidates, that would need to offer advantages compared to a simulated Boson sampling device.
Our techniques can be used to translate Boson sampling devices onto digital quantum computers, hence it naturally enables those kind of simulations.
To illustrate this, we transform one of the first experimental Boson sampling setups~\cite{crespi2014} using integrated photonics into the language of digital quantum computers and reproduce its results.
The corresponding setup is shown in Fig.~\ref{fig:crespi_example}. We simulate it by representing each path with 2 qubits, resulting in a 10-qubit quantum circuit. The corresponding distribution, using different Trotter numbers, is shown in Fig.~\ref{fig:crespi_example}, where distributions obtained with 40-50 trotter steps are visually indistinguishable from the theoretically-obtained distribution in Ref.~\cite{crespi2014}, and the low order simulation with 5 Trotter steps already produces a qualitatively correct result that agrees with the results of Ref.~\cite{crespi2014}.
We used the number of \textit{valid states} - obeying particle number conservation - as a potential indicator for sufficient accuracy in the Beam-Splitter representation.
We note here that there is still significant potential for further improvement with respect to the explicit construction of the Trotter expansion, for example with randomized compilation~\cite{childs2019random} or by optimizing the ordering in the Trotter decomposition using similar ideas as those used in Fermionic simulation (see Refs.~\cite{grimsley2020} and~\cite{izmaylov2020order}). 
In our opinion, the digital quantum simulation of a photonic Boson sampling setup will not bring any advantages regarding sample sizes and runtime. It is furthermore a tool that might be applied to analyze the evolution of the wavefunction in the same way as it is currently done with classical simulators. Our approach aims to take over, at a point where explicit classical simulation is no longer feasible. Other than the optimization example of the last section, our Boson sampling example is unlikely to be executable on recently emerging architectures due to the gate counts in the quantum circuits. Due to it's conceptual simplicity and low qubit requirements we think it could be an interesting benchmark example for early stage fault tolerant machines and advanced compilation strategies.

\section{Conclusions}
In 1981, Richard Feynman gave his visionary keynote speech at MIT that paved the way towards quantum simulations~\cite{feynman1982simulating}. He explained his intuitions using an quantum optical Einstein-Podolsky-Rosen experiment and famously concludes that \textit{nature isn't classical, dammit, and if you want to make a simulation of nature, you'd better make it quantum mechanical}.
As a direct consequence of Feynman's insight, we argue that the very hardware for simulating and measuring nature isn't classical, and thereby \textit{you'd better make} design and verification \textit{quantum mechanical}.
Here, we have shown quantum simulation of quantum hardware for entangled quantum photonic systems -- the basis of Feynman's original thought experiment. However, the general idea that we propose here goes far beyond that. Any hardware that measures, transforms or exploits quantum systems should ultimately be designed through methods that leverage the power of quantum computers to exploit the full potential furnished by quantum physics~\cite{kyaw2020}. Combinations with classical algorithms should however not be restricted but rather leveraged to gain the best of both worlds. {Compared to working directly with the photonic setups, our proposal brings several advantages. First, it allows the usage of operations that are often easy to realize on a digital quantum computer, such as the direct preparation of the target and initial states, allowing for less measurement intensive optimization protocols and avoiding long waiting times associated to the low count rates of the initial photon creation. Second, the formulation within a digital quantum computing framework gives direct access of the associated gradients through decompositions into directly differentiable quantum gates.~\cite{schuld2019, kottmann2020feasible, tequila} Finally, the abstract formulation within a generalized, automatically differentiable physical language, in the form of quantum circuits and expectation values, make it independent of special purpose simulation techniques. Our open-source implementation of the techniques developed here will profit immediately from improvements in quantum hardware as well as classical simulators.} In future applications, the techniques for optimization and encoding developed in this work could be used as an efficient sub-routine of graph based topological optimization algorithms~\cite{krenn2020conceptual}. As such, we anticipate the applications of \textit{quantum designed quantum hardware} to quantum computing hardware, quantum sensors, quantum memories, and quantum communication networks.\\

Python code for the \textsc{tequila}~\cite{tequila} package as well as explicit code~\cite{code-repo} for the calculations in this work can be found on github.

\section*{Acknowledgements}
A.A.-G. acknowledges the generous support from Google, Inc.  in the form of a Google Focused Award. A.A.-G. also acknowledges support from the Canada Industrial Research Chairs  Program and the Canada 150 Research Chairs Program. We acknowledge the support of the U.S. Department of Energy through grant \#DE-AC02-05CH11231 subgrant LBNL - \#505736.
We thank the generous support of Anders G. Fr\o{}seth. MK acknowledges support from the FWF via the Erwin Schr\"odinger fellowship No. J4309. 

\clearpage
\section*{References}
\bibliographystyle{iopart-num}
\bibliography{refs.bib}

\clearpage
\appendix

\section{Binary Encoding}\label{sec:appendix_binary_encoding}
\begin{figure}[h]
    \centering
    \savebox{\imagebox}{
 		 \input{pics/qubits.tex}
 	}
    
    \subfigure[]{\label{fig:mapping-paths}
        \centering\raisebox{\dimexpr.5\ht\imagebox-.5\height}{
            \input{pics/path.tex}
        }
    }
    \subfigure[]{\label{fig:mapping-modes}
        \centering\raisebox{\dimexpr.5\ht\imagebox-.5\height}{
            \input{pics/modes.tex}
        }
    }
    \subfigure[]{\label{fig:mapping-qubits}
    \centering\usebox{\imagebox}
    }
    \caption{Mapping of a photonic setup into qubits. (a) Optical path. (b) Internal degrees of freedom. (c) Qubits representing the photon occupation number of each internal degree of freedom [0-7].}
    \label{fig:mapping}
\end{figure}
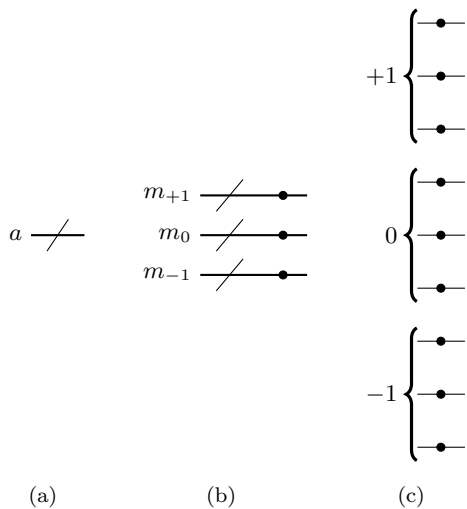

In the binary encoding, the usual bosonic operators are represented as
\begin{equation}
	\hat{a}^\dagger = \sum_{n=0}^{d-2} \sqrt{n+1} \ketbra{n+1}{n}.
\end{equation}
The operators $ \ketbra{n+1}{n}$ are then mapped to strings of Pauli operators, after converting $n=a_0 2^k + a_1 2^{k-1}+\cdots+a_{k-1}2^1 + a_{k}  2^0$ into binary and using the following relations:
\begin{eqnarray}
	\ketbra{0}{1} &= (X+i Y)/2,\quad
	\ketbra{0}{0} = (\mathbb{I}+Z)/2,\\
	\ketbra{1}{0} &= (X-i Y)/2,\quad
	\ketbra{1}{1} = (\mathbb{I}-Z)/2.
\end{eqnarray}
where, $X, Y, Z$ are the usual Pauli matrices and $\mathbb{I}$ is the identity.
Fig.~\ref{fig:mapping} illustrates the mapping of a photonic path $a$ with internal degrees of freedom (labeled as -1, 0, +1) and where each internal degree of freedom is represented by multiple qubits encoding the photonic occupation number in binary.

\section{Optical elements}
\subsection{Beam Splitter ($BS$)}\label{sec:appendix_beam_splitter}
\begin{figure}[h]
    \centering
    \savebox{\imagebox}{\input{pics/bs_abstract.tex}}
    \subfigure[]{\label{fig:bs_abstract}
        \centering\usebox{\imagebox}
    }
    \subfigure[]{\label{fig:bs_qudit}
        \centering\raisebox{\dimexpr.5\ht\imagebox-.5\height}{
            \input{pics/bs_qudit.tex}
        }
    }
    \caption{Beam splitter acting two photonic paths with three internal modes each. \ref{fig:bs_abstract} abstract representation. \ref{fig:bs_qudit} More detailed representation with individual unitaries acting on the different internal modes}
    \label{fig:bs}
\end{figure}
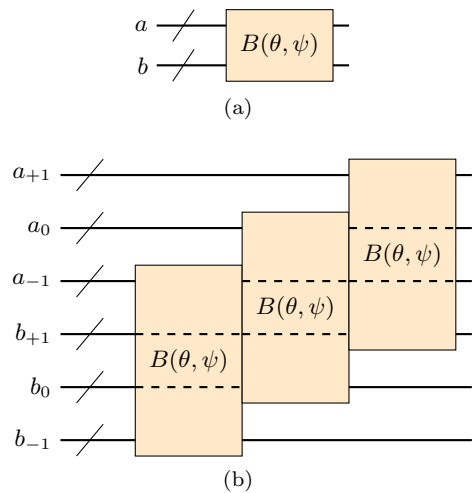

A beam splitter which acts on two photonic paths $a$ and $b$ can be described as
\begin{eqnarray}
    BS\lr{\theta, \psi} &= e^{i\theta\sum_m\lr{\psi\cop{a}_m\aop{b}_m + \psi^*\cop{b}_m\aop{a}_m}} \nonumber\\ 
    &= \prod_m e^{i\theta\lr{\psi\cop{a}_m\aop{b}_m + \psi^*\cop{b}_m\aop{a}_m}}\label{eq:bs}.
\end{eqnarray}
where $\theta$ and $\psi$ are complex numbers and the operator $\cop{a}_m$ creates a photon in internal mode $m$ in the photonic path $a$.
Since each term in the sum in the exponent commutes, the part which depends on the internal modes separates naturally into a product of unitaries acting only on a specific mode in each path (note that we formally apply the qubit encoding after this step).
This is illustrated in Fig.~\ref{fig:bs}.
The Trotter decomposition (used for example in Fig.~\ref{fig:crespi_example} of the main text) becomes necessary after the photonic operators have been mapped onto Pauli operators, \textit{i.e} the parts acting on the individual modes $m$ are trotterized after they are mapped onto qubit operators. 

\subsection{Phase shifter ($P$)}\label{sec:appendix_phase_shifter}
\begin{figure}[h]
    \centering
    \savebox{\imagebox}{\input{pics/ps_abstract.tex}}
    \subfigure[]{\label{fig:ps_qudit}
        \centering\usebox{\imagebox}
    }
    \subfigure[]{\label{fig:ps_qubit}
        \centering\raisebox{\dimexpr.5\ht\imagebox-.5\height}{
            \input{pics/ps_qubit.tex}
        }
    }
    \caption{Phase shifter acting on path $a$ with one internal mode. \ref{fig:ps_qudit} abstract representation. \ref{fig:ps_qubit} Explicit representation with 3 qubits per mode.}
    \label{fig:ps}
\end{figure}
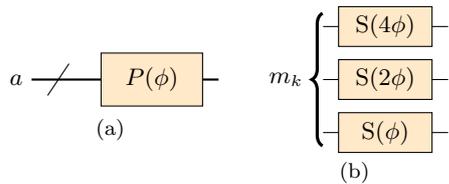

The phase shifter can be defined by its action onto a single internal degree of freedom in a given photonic path.
It acts by adding a phase $\phi$ to each photonic state w.r.t the number of photons occupying it.
In terms of creation/annihilation operators this can be expressed as 
\begin{eqnarray}
    P\lr{\phi} = e^{i\phi \cop{a} \aop{a}}.
\end{eqnarray}
where we have assumed only one internal degree of freedom in the path.
In binary representation the action of the phase shifter can be implemented by a set of single qubit phase gates $S\lr{\phi}$ which add a phase $e^{i\phi}$ to the $\ket{1}$ state and acts trivial on $\ket{0}$.
Assume for example that the internal mode is represented by $n$ qubits and that the occupation number is encoded in binary, then the phase shifter acting on that mode can be implemented as
\begin{eqnarray}
    P\lr{\phi} \xrightarrow{n-qubits} S\lr{2^{n-1}\phi} \otimes S\lr{2^{n-2}\phi} \otimes \dots S\lr{\phi}
\end{eqnarray}
where we assumed most significant ordering in the binary encoding.
This implementation of the phase shifter with phase gates is also shown in Fig.~\ref{fig:ps}.
Note that each $S\lr{\phi}$ gate can be replaced by a $R_z\lr{-\phi}$ rotation which implements the same relative phase. 

\subsection{Mode-dependent phase shifter, Dove prism ($DP$)}
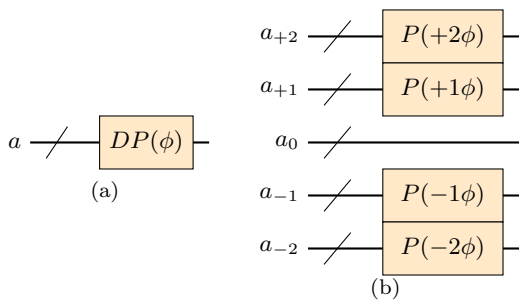
\begin{figure}[h]
    \centering
    \savebox{\imagebox}{\input{pics/dp_abstract.tex}}
    \subfigure[]{\label{fig:dp_abstract}
        \usebox{\imagebox}
    }
    \subfigure[]{\label{fig:dp_qubit}
        \centering\raisebox{\dimexpr0.5\ht\imagebox-0.5\height}{
             \input{pics/dp_qudit.tex}
        }
    }
    \caption{Mode dependent phase shifter (Dove prism) acting on path $a$ with five internal modes. \ref{fig:dp_abstract} abstract representation. \ref{fig:dp_qubit} More detailed representation showing the internal modes. For the qubit-wise representation of each phase shifter see Fig.~\ref{fig:ps}}
    \label{fig:dp}
\end{figure}

If internal degrees of freedom are used within the photonic paths a mode dependent phase shifter (Dove prism) can be applied.
In the case of orbital angular momentum as inner degree of freedem the Dove prism acts like a set of mode dependent phase shifters
\begin{eqnarray}
    DP\lr{\phi} = \prod_{m \in \left\{ \dots +1, 0, -1 \dots \right\}} P\lr{m\phi}
\end{eqnarray}
where $m$ is the orbital angular momentum quantum number.
We illustrate the implementation of a Dove prism in Fig.~\ref{fig:dp}

\subsection{Photonic swap, Hologram and Mirror}\label{sec:appendix_photonic_swap_hologram_mirror}
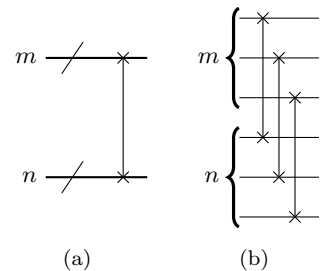
\begin{figure}[h]
    \centering
    \savebox{\imagebox}{\input{pics/swap_qudit.qpic.tex}}
    \subfigure[]{\label{fig:photonic_swap_qudit}
        \usebox{\imagebox}
    }
    \subfigure[]{\label{fig:photonic_swap_qubit}
        \raisebox{\dimexpr0.5\ht\imagebox-0.5\height}{
             \input{pics/swap_qubit.qpic.tex}
        }
    }
    \caption{Swap gate between two photonic modes $m$ and $n$ which can be in the same path or in different paths. \ref{fig:photonic_swap_qudit} abstract representation. \ref{fig:photonic_swap_qubit} Explicit representation with 3 qubits per mode.}
    \label{fig:photonic_swap}
\end{figure}

In the following we show the implementation of two non-parametrized optical elements which are called hologram and mirror and both depend on photonic swap operations which swap photons between two photonic modes or paths.
The implementation of those swap operations is straight forward by applying a swap operation onto all qubits which represent the photonic mode or path.
We illustrate this in Fig.~\ref{fig:photonic_swap}.
In the main text we used the compressed notation of the photonic swaps in Fig.~\ref{fig:overview}:.

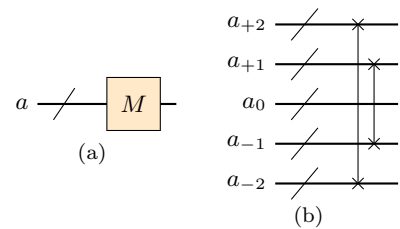
\begin{figure}[h]
    \centering
    \savebox{\imagebox}{\input{pics/mirror_abstract.tex}}
    \subfigure[]{\label{fig:mirror_abstract}
        \usebox{\imagebox}
    }
    \subfigure[]{\label{fig:mirror_qudit}
        \raisebox{\dimexpr0.5\ht\imagebox-0.5\height}{
            \input{pics/mirror.qpic.tex}
        }
    }
    \caption{Mirror acting on internal photonic modes in a given path. \ref{fig:mirror_qudit} abstract representation. \ref{fig:mirror_qudit} Explicit representation with photonic swap gates (see Fig.~\ref{fig:photonic_swap})}
    \label{fig:mirror}
\end{figure}

The mirror acts on the internal degrees of freedom in a single photonic path by changing the sign of the internal degrees of freedom.
Assume for example three internal modes and a state with $n_i$ photons in each internal mode $i \in \left\{+1,0,-1\right\}$, then the operation which represents the mirror element acts as
\begin{eqnarray}
    M\ket{l_{-1},m_0,n_1} = \ket{n_{-1},m_0,l_{1}}.\label{eq:mirror_example}
\end{eqnarray}
The mirror can be implemented straight forwardly with the swap operations defined in Fig.~\ref{fig:photonic_swap}.
In Fig.~\ref{fig:mirror} this is illustrated.

\begin{figure}[h]
    \centering
    \savebox{\imagebox}{\input{pics/hologram_abstract.tex}}
    \subfigure[]{\label{fig:hologram_abstract}
        \usebox{\imagebox}
    }
    \subfigure[]{\label{fig:hologram_qudit}
    \raisebox{\dimexpr0.5\ht\imagebox-0.5\height}{
         \input{pics/hologram.qpic.tex}
    }
    }
    \caption{Cyclic approximation of a hologram acting two photonic paths with five internal modes each. \ref{fig:hologram_abstract} abstract representation. \ref{fig:hologram_qudit}Explicit representation with photonic swap gates (see Fig.~\ref{fig:photonic_swap})}
    \label{fig:hologram}
\end{figure}
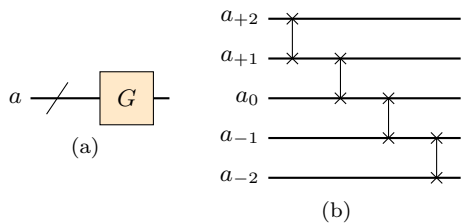

The hologram ($G$) acts on the internal degrees of freedom by increasing them by one.
Take for example the state with $l, m, n$ photons in modes $-1, 0, 1$, then the hologram will act as
\begin{eqnarray}
    G\ket{l_{-1},m_{0},n_{+1}} = \ket{0_{-1}, l_{0}, m_{+1}, n_{+2}},
\end{eqnarray}
resulting state with no photons in mode -1, $l$ photons in mode 0, $m$ photons in mode 1 and $n$ photons in mode 2 which didn't had any photons before.
The action of the hologram is clearly not unitary for a truncated number of internal degrees of freedom.
In order to implement its action we propose a cyclic version shown in Fig.~\ref{fig:hologram}. For internal degrees of freedom from $-m$ to $m$, this acts as: 
\begin{eqnarray}
	&\tilde{G}\ket{n_{-m}, \dots, n_{0}, \dots, n_{m}} \nonumber\\&= \ket{n_{m}, n_{-m}, \dots, n_{1}, \dots, n_{m-1} }.
\end{eqnarray}
Simulating optical setups with the cyclic hologram can be performed as long as the number of represented internal degrees of freedom is not restricted too much, meaning that the cutoff has to be carefully chosen in a way that the represented highest internal degree of freedom is never occupied in the underlying setup or at least that its occupation is unlikely.\\

The circuits we show here assume binary mapping when refined until the qubit level (see Sec.~\ref{sec:circuits} and Ref.~\cite{nicolas2019}).
A generalization to unary and Gray code can be achieved in a straight forward way by permuting the individual gates on the qubit level for the phase shifter and for the beam splitter by mapping the creation/annihilation operators in the same way as described in Ref.~\cite{nicolas2019}. The implementation of the swap gate, mirror and hologram will stay the same for all three mappings.\\

Neither hologram nor mirror are used explicitly in the computations of this work.
They are however important to extend the \textit{332-state} preparation setup to prepare a multidimensional \textit{GHZ} state (see Fig.~2a of Ref.~\cite{krenn2016automated}).

\section{Implementation of General Elements}

General elements can be represented in similar ways as the beam splitter by mapping their second-quantized generators onto qubits followed by a decomposition of the generated unitary into sequences of primitive gates.
Depending on the element, a customized approach (such as given here for the hologram, mirror and phase shifter) might however lead to more optimized gate sequences for the specific element.

In addition to the implementation of concrete optical elements in a gate-based QComputer, we also indicate in an example how more abstract representations of quantum optics can be mapped to circuits. Here we will use an abstract \textit{edge} within a generalized graph-based representation of quantum optical setups developed in~\cite{krenn2020conceptual}.
Those graph representation can, for example, be translated into non-linear crystals that create photon pairs in specific modes (SPDC) or on-chip four-wave mixing processes (SFWM).  
Consider an \textit{edge} of Ref.~\cite{krenn2020conceptual} connecting vertices (photonic paths) $a$ and $b$ and having colors (photonic modes) $m$ and $n$. The generator $G$ that generates the unitary of the edge $U(\omega)=e^{-i\frac{\omega}{2}G}$ is
\begin{eqnarray}
    G_{a_m b_n} = a^\dagger_{m}b^\dagger_{n} - a_{m}b_{n} .
\end{eqnarray}
With the mappings of~\cite{nicolas2019} this generator can be mapped to Pauli strings and the generated unitary can be decomposed via a Trotter decomposition. One of the proposals in the main text, is to combine those graph based representations with the techniques of this work as the simulation of the graph (currently achieved over explicit series expansion of the unitary, neglecting the hermitian adjoint in the generator) is currently the computational bottleneck of the \texttt{theseus} algorithm developed in~\cite{krenn2020conceptual}. For further illustration, we added an additional example on github~\cite{code-repo} that illustrates the optimization of a GHZ state via a small graph based setup within our implementation.

\twocolumngrid

\end{document}

%% file: pics/qubits.tex
%! \usetikzlibrary{decorations.pathreplacing,decorations.pathmorphing}
\begin{tikzpicture}[scale=1.000000,x=1pt,y=1pt]
\filldraw[color=white] (0.000000, -10.000000) rectangle (18.000000, 170.000000);
% Drawing wires
% Line 2: a b c W {+1}< breadth=20
\draw[color=black] (0.000000,160.000000) -- (18.000000,160.000000);
%   Deferring wire label at (0.000000,160.000000)
% Line 2: a b c W {+1}< breadth=20
\draw[color=black] (0.000000,140.000000) -- (18.000000,140.000000);
%   Deferring wire label at (0.000000,140.000000)
% Line 2: a b c W {+1}< breadth=20
\draw[color=black] (0.000000,120.000000) -- (18.000000,120.000000);
\filldraw[color=white,fill=white] (0.000000,115.000000) rectangle (-4.000000,165.000000);
\draw[decorate,decoration={brace,amplitude = 4.000000pt},very thick] (0.000000,115.000000) -- (0.000000,165.000000);
\draw[color=black] (-4.000000,140.000000) node[left] {${+1}$};
% Line 3: d e f W {0}< breadth=20
\draw[color=black] (0.000000,100.000000) -- (18.000000,100.000000);
%   Deferring wire label at (0.000000,100.000000)
% Line 3: d e f W {0}< breadth=20
\draw[color=black] (0.000000,80.000000) -- (18.000000,80.000000);
%   Deferring wire label at (0.000000,80.000000)
% Line 3: d e f W {0}< breadth=20
\draw[color=black] (0.000000,60.000000) -- (18.000000,60.000000);
\filldraw[color=white,fill=white] (0.000000,55.000000) rectangle (-4.000000,105.000000);
\draw[decorate,decoration={brace,amplitude = 4.000000pt},very thick] (0.000000,55.000000) -- (0.000000,105.000000);
\draw[color=black] (-4.000000,80.000000) node[left] {${0}$};
% Line 4: g h i W {-1}< breadth=20
\draw[color=black] (0.000000,40.000000) -- (18.000000,40.000000);
%   Deferring wire label at (0.000000,40.000000)
% Line 4: g h i W {-1}< breadth=20
\draw[color=black] (0.000000,20.000000) -- (18.000000,20.000000);
%   Deferring wire label at (0.000000,20.000000)
% Line 4: g h i W {-1}< breadth=20
\draw[color=black] (0.000000,0.000000) -- (18.000000,0.000000);
\filldraw[color=white,fill=white] (0.000000,-5.000000) rectangle (-4.000000,45.000000);
\draw[decorate,decoration={brace,amplitude = 4.000000pt},very thick] (0.000000,-5.000000) -- (0.000000,45.000000);
\draw[color=black] (-4.000000,20.000000) node[left] {${-1}$};
% Done with wires; drawing gates
% Line 6: a
\filldraw (9.000000, 160.000000) circle(1.500000pt);
% Line 7: b
\filldraw (9.000000, 140.000000) circle(1.500000pt);
% Line 8: c
\filldraw (9.000000, 120.000000) circle(1.500000pt);
% Line 9: d
\filldraw (9.000000, 100.000000) circle(1.500000pt);
% Line 10: e
\filldraw (9.000000, 80.000000) circle(1.500000pt);
% Line 11: f
\filldraw (9.000000, 60.000000) circle(1.500000pt);
% Line 12: g
\filldraw (9.000000, 40.000000) circle(1.500000pt);
% Line 13: h
\filldraw (9.000000, 20.000000) circle(1.500000pt);
% Line 14: i
\filldraw (9.000000, 0.000000) circle(1.500000pt);
% Done with gates; drawing ending labels
% Done with ending labels; drawing cut lines and comments
% Done with comments
\end{tikzpicture}

%% file: pics/path.tex
\begin{tikzpicture}[scale=1.000000,x=1pt,y=1pt]
\filldraw[color=white] (0.000000, -7.500000) rectangle (20.000000, 7.500000);
% Drawing wires
% Line 1: style=thick a W a
\draw[color=black,thick] (0.000000,0.000000) -- (20.000000,0.000000);
\draw[color=black] (0.000000,0.000000) node[left] {$a$};
% Done with wires; drawing gates
% Line 3: a /
\draw (6.000000, -6.000000) -- (14.000000, 6.000000);
% Done with gates; drawing ending labels
% Done with ending labels; drawing cut lines and comments
% Done with comments
\end{tikzpicture}

%% file: pics/modes.tex
\begin{tikzpicture}[scale=1.000000,x=1pt,y=1pt]
\filldraw[color=white] (0.000000, -7.500000) rectangle (40.000000, 37.500000);
% Drawing wires
% Line 1: style=thick b W {m_{+1}}
\draw[color=black,thick] (0.000000,30.000000) -- (40.000000,30.000000);
\draw[color=black] (0.000000,30.000000) node[left] {${m_{+1}}$};
% Line 2: style=thick c W {m_{ 0}}
\draw[color=black,thick] (0.000000,15.000000) -- (40.000000,15.000000);
\draw[color=black] (0.000000,15.000000) node[left] {${m_{ 0}}$};
% Line 3: style=thick d W {m_{-1}}
\draw[color=black,thick] (0.000000,0.000000) -- (40.000000,0.000000);
\draw[color=black] (0.000000,0.000000) node[left] {${m_{-1}}$};
% Done with wires; drawing gates
% Line 5: b c d  / width=10
\draw (6.000000, 24.000000) -- (16.000000, 36.000000);
\draw (6.000000, 9.000000) -- (16.000000, 21.000000);
\draw (6.000000, -6.000000) -- (16.000000, 6.000000);
% Line 7: b
\filldraw (31.000000, 30.000000) circle(1.500000pt);
% Line 8: c
\filldraw (31.000000, 15.000000) circle(1.500000pt);
% Line 9: d
\filldraw (31.000000, 0.000000) circle(1.500000pt);
% Done with gates; drawing ending labels
% Done with ending labels; drawing cut lines and comments
% Done with comments
\end{tikzpicture}

%% file: pics/bs_abstract.tex
\begin{tikzpicture}[scale=1.000000,x=1pt,y=1pt]
\filldraw[color=white] (0.000000, -7.500000) rectangle (72.000000, 22.500000);
% Drawing wires
% Line 1: style=thick a W a
\draw[color=black,thick] (0.000000,15.000000) -- (72.000000,15.000000);
\draw[color=black] (0.000000,15.000000) node[left] {$a$};
% Line 2: style=thick b W b
\draw[color=black,thick] (0.000000,0.000000) -- (72.000000,0.000000);
\draw[color=black] (0.000000,0.000000) node[left] {$b$};
% Done with wires; drawing gates
% Line 4: a /
\draw (6.000000, 9.000000) -- (14.000000, 21.000000);
% Line 5: b /
\draw (6.000000, -6.000000) -- (14.000000, 6.000000);
% Line 7: a b G $B(\theta, \psi)$ width=40 height=20 fill=red!70!white!50!yellow!25!
\draw (46.000000,15.000000) -- (46.000000,0.000000);
\begin{scope}
\draw[fill=red!70!white!50!yellow!25!] (46.000000, 7.500000) +(-45.000000:28.284271pt and 19.091883pt) -- +(45.000000:28.284271pt and 19.091883pt) -- +(135.000000:28.284271pt and 19.091883pt) -- +(225.000000:28.284271pt and 19.091883pt) -- cycle;
\clip (46.000000, 7.500000) +(-45.000000:28.284271pt and 19.091883pt) -- +(45.000000:28.284271pt and 19.091883pt) -- +(135.000000:28.284271pt and 19.091883pt) -- +(225.000000:28.284271pt and 19.091883pt) -- cycle;
\draw (46.000000, 7.500000) node {$B(\theta, \psi)$};
\end{scope}
% Done with gates; drawing ending labels
% Done with ending labels; drawing cut lines and comments
% Done with comments
\end{tikzpicture}

%% file: pics/bs_qudit.tex
\begin{tikzpicture}[scale=1.000000,x=1pt,y=1pt]
\filldraw[color=white] (0.000000, -10.000000) rectangle (154.000000, 110.000000);
% Drawing wires
% Line 1: style=thick a0 W {a_{+1}} breadth=20
\draw[color=black,thick] (0.000000,100.000000) -- (154.000000,100.000000);
\draw[color=black] (0.000000,100.000000) node[left] {${a_{+1}}$};
% Line 2: style=thick b0 W {a_{ 0}} breadth=20
\draw[color=black,thick] (0.000000,80.000000) -- (154.000000,80.000000);
\draw[color=black] (0.000000,80.000000) node[left] {${a_{ 0}}$};
% Line 3: style=thick c0 W {a_{-1}} breadth=20
\draw[color=black,thick] (0.000000,60.000000) -- (154.000000,60.000000);
\draw[color=black] (0.000000,60.000000) node[left] {${a_{-1}}$};
% Line 5: style=thick a1 W {b_{+1}} breadth=20
\draw[color=black,thick] (0.000000,40.000000) -- (154.000000,40.000000);
\draw[color=black] (0.000000,40.000000) node[left] {${b_{+1}}$};
% Line 6: style=thick b1 W {b_{ 0}} breadth=20
\draw[color=black,thick] (0.000000,20.000000) -- (154.000000,20.000000);
\draw[color=black] (0.000000,20.000000) node[left] {${b_{ 0}}$};
% Line 7: style=thick c1 W {b_{-1}} breadth=20
\draw[color=black,thick] (0.000000,0.000000) -- (154.000000,0.000000);
\draw[color=black] (0.000000,0.000000) node[left] {${b_{-1}}$};
% Done with wires; drawing gates
% Line 11: a0 b0 c0 a1 b1 c1 / width=10
\draw (6.000000, 94.000000) -- (16.000000, 106.000000);
\draw (6.000000, 74.000000) -- (16.000000, 86.000000);
\draw (6.000000, 54.000000) -- (16.000000, 66.000000);
\draw (6.000000, 34.000000) -- (16.000000, 46.000000);
\draw (6.000000, 14.000000) -- (16.000000, 26.000000);
\draw (6.000000, -6.000000) -- (16.000000, 6.000000);
% Line 13: c0 c1 G {$B(\theta, \psi)$} width=40 height=15 fill=red!70!white!50!yellow!25!
\draw (48.000000,60.000000) -- (48.000000,0.000000);
\begin{scope}
\draw[fill=red!70!white!50!yellow!25!] (48.000000, 30.000000) +(-45.000000:28.284271pt and 50.911688pt) -- +(45.000000:28.284271pt and 50.911688pt) -- +(135.000000:28.284271pt and 50.911688pt) -- +(225.000000:28.284271pt and 50.911688pt) -- cycle;
\clip (48.000000, 30.000000) +(-45.000000:28.284271pt and 50.911688pt) -- +(45.000000:28.284271pt and 50.911688pt) -- +(135.000000:28.284271pt and 50.911688pt) -- +(225.000000:28.284271pt and 50.911688pt) -- cycle;
\draw (48.000000, 30.000000) node {{$B(\theta, \psi)$}};
\end{scope}
\draw[color=black,dashed,thick] (28.000000, 40.000000) -- (68.000000, 40.000000);
\draw[color=black,dashed,thick] (28.000000, 20.000000) -- (68.000000, 20.000000);
% Line 14: b0 b1 G {$B(\theta, \psi)$} width=40 height=15 fill=red!70!white!50!yellow!25!
\draw (88.000000,80.000000) -- (88.000000,20.000000);
\begin{scope}
\draw[fill=red!70!white!50!yellow!25!] (88.000000, 50.000000) +(-45.000000:28.284271pt and 50.911688pt) -- +(45.000000:28.284271pt and 50.911688pt) -- +(135.000000:28.284271pt and 50.911688pt) -- +(225.000000:28.284271pt and 50.911688pt) -- cycle;
\clip (88.000000, 50.000000) +(-45.000000:28.284271pt and 50.911688pt) -- +(45.000000:28.284271pt and 50.911688pt) -- +(135.000000:28.284271pt and 50.911688pt) -- +(225.000000:28.284271pt and 50.911688pt) -- cycle;
\draw (88.000000, 50.000000) node {{$B(\theta, \psi)$}};
\end{scope}
\draw[color=black,dashed,thick] (68.000000, 60.000000) -- (108.000000, 60.000000);
\draw[color=black,dashed,thick] (68.000000, 40.000000) -- (108.000000, 40.000000);
% Line 15: a0 a1 G {$B(\theta, \psi)$} width=40 height=15 fill=red!70!white!50!yellow!25!
\draw (128.000000,100.000000) -- (128.000000,40.000000);
\begin{scope}
\draw[fill=red!70!white!50!yellow!25!] (128.000000, 70.000000) +(-45.000000:28.284271pt and 50.911688pt) -- +(45.000000:28.284271pt and 50.911688pt) -- +(135.000000:28.284271pt and 50.911688pt) -- +(225.000000:28.284271pt and 50.911688pt) -- cycle;
\clip (128.000000, 70.000000) +(-45.000000:28.284271pt and 50.911688pt) -- +(45.000000:28.284271pt and 50.911688pt) -- +(135.000000:28.284271pt and 50.911688pt) -- +(225.000000:28.284271pt and 50.911688pt) -- cycle;
\draw (128.000000, 70.000000) node {{$B(\theta, \psi)$}};
\end{scope}
\draw[color=black,dashed,thick] (108.000000, 80.000000) -- (148.000000, 80.000000);
\draw[color=black,dashed,thick] (108.000000, 60.000000) -- (148.000000, 60.000000);
% Done with gates; drawing ending labels
% Done with ending labels; drawing cut lines and comments
% Done with comments
\end{tikzpicture}

%% file: pics/ps_abstract.tex
\begin{tikzpicture}[scale=1.000000,x=1pt,y=1pt]
\filldraw[color=white] (0.000000, -7.500000) rectangle (70.000000, 7.500000);
% Drawing wires
% Line 1: style=thick a W a
\draw[color=black,thick] (0.000000,0.000000) -- (70.000000,0.000000);
\draw[color=black] (0.000000,0.000000) node[left] {$a$};
% Done with wires; drawing gates
% Line 3: a /
\draw (6.000000, -6.000000) -- (14.000000, 6.000000);
% Line 5: a G $P(\phi)$ width=38 height=20 fill=red!70!white!50!yellow!25!
\begin{scope}
\draw[fill=red!70!white!50!yellow!25!] (45.000000, -0.000000) +(-45.000000:26.870058pt and 14.142136pt) -- +(45.000000:26.870058pt and 14.142136pt) -- +(135.000000:26.870058pt and 14.142136pt) -- +(225.000000:26.870058pt and 14.142136pt) -- cycle;
\clip (45.000000, -0.000000) +(-45.000000:26.870058pt and 14.142136pt) -- +(45.000000:26.870058pt and 14.142136pt) -- +(135.000000:26.870058pt and 14.142136pt) -- +(225.000000:26.870058pt and 14.142136pt) -- cycle;
\draw (45.000000, -0.000000) node {$P(\phi)$};
\end{scope}
% Done with gates; drawing ending labels
% Done with ending labels; drawing cut lines and comments
% Done with comments
\end{tikzpicture}

%% file: pics/ps_qubit.tex
%! \usetikzlibrary{decorations.pathreplacing,decorations.pathmorphing}
\begin{tikzpicture}[scale=1.000000,x=1pt,y=1pt]
\filldraw[color=white] (0.000000, -10.000000) rectangle (47.000000, 50.000000);
% Drawing wires
% Line 1: a b c W {m_k}< breadth=20
\draw[color=black] (0.000000,40.000000) -- (47.000000,40.000000);
%   Deferring wire label at (0.000000,40.000000)
% Line 1: a b c W {m_k}< breadth=20
\draw[color=black] (0.000000,20.000000) -- (47.000000,20.000000);
%   Deferring wire label at (0.000000,20.000000)
% Line 1: a b c W {m_k}< breadth=20
\draw[color=black] (0.000000,0.000000) -- (47.000000,0.000000);
\filldraw[color=white,fill=white] (0.000000,-5.000000) rectangle (-4.000000,45.000000);
\draw[decorate,decoration={brace,amplitude = 4.000000pt},very thick] (0.000000,-5.000000) -- (0.000000,45.000000);
\draw[color=black] (-4.000000,20.000000) node[left] {${m_k}$};
% Done with wires; drawing gates
% Line 3: c G {$\text{S}(\phi)$} height=15 width=35 fill=red!70!white!50!yellow!25!
\begin{scope}
\draw[fill=red!70!white!50!yellow!25!] (23.500000, -0.000000) +(-45.000000:24.748737pt and 10.606602pt) -- +(45.000000:24.748737pt and 10.606602pt) -- +(135.000000:24.748737pt and 10.606602pt) -- +(225.000000:24.748737pt and 10.606602pt) -- cycle;
\clip (23.500000, -0.000000) +(-45.000000:24.748737pt and 10.606602pt) -- +(45.000000:24.748737pt and 10.606602pt) -- +(135.000000:24.748737pt and 10.606602pt) -- +(225.000000:24.748737pt and 10.606602pt) -- cycle;
\draw (23.500000, -0.000000) node {{$\text{S}(\phi)$}};
\end{scope}
% Line 4: b G {$\text{S}(2\phi)$} height=15 width=35 fill=red!70!white!50!yellow!25!
\begin{scope}
\draw[fill=red!70!white!50!yellow!25!] (23.500000, 20.000000) +(-45.000000:24.748737pt and 10.606602pt) -- +(45.000000:24.748737pt and 10.606602pt) -- +(135.000000:24.748737pt and 10.606602pt) -- +(225.000000:24.748737pt and 10.606602pt) -- cycle;
\clip (23.500000, 20.000000) +(-45.000000:24.748737pt and 10.606602pt) -- +(45.000000:24.748737pt and 10.606602pt) -- +(135.000000:24.748737pt and 10.606602pt) -- +(225.000000:24.748737pt and 10.606602pt) -- cycle;
\draw (23.500000, 20.000000) node {{$\text{S}(2\phi)$}};
\end{scope}
% Line 5: a G {$\text{S}(4\phi)$} height=15 width=35 fill=red!70!white!50!yellow!25!
\begin{scope}
\draw[fill=red!70!white!50!yellow!25!] (23.500000, 40.000000) +(-45.000000:24.748737pt and 10.606602pt) -- +(45.000000:24.748737pt and 10.606602pt) -- +(135.000000:24.748737pt and 10.606602pt) -- +(225.000000:24.748737pt and 10.606602pt) -- cycle;
\clip (23.500000, 40.000000) +(-45.000000:24.748737pt and 10.606602pt) -- +(45.000000:24.748737pt and 10.606602pt) -- +(135.000000:24.748737pt and 10.606602pt) -- +(225.000000:24.748737pt and 10.606602pt) -- cycle;
\draw (23.500000, 40.000000) node {{$\text{S}(4\phi)$}};
\end{scope}
% Done with gates; drawing ending labels
% Done with ending labels; drawing cut lines and comments
% Done with comments
\end{tikzpicture}

%% file: pics/dp_abstract.tex
\begin{tikzpicture}[scale=1.000000,x=1pt,y=1pt]
\filldraw[color=white] (0.000000, -7.500000) rectangle (67.000000, 7.500000);
% Drawing wires
% Line 1: style=thick a W a
\draw[color=black,thick] (0.000000,0.000000) -- (67.000000,0.000000);
\draw[color=black] (0.000000,0.000000) node[left] {$a$};
% Done with wires; drawing gates
% Line 3: a /
\draw (6.000000, -6.000000) -- (14.000000, 6.000000);
% Line 5: a G $DP(\phi)$ width=35 height=20 fill=red!70!white!50!yellow!25!
\begin{scope}
\draw[fill=red!70!white!50!yellow!25!] (43.500000, -0.000000) +(-45.000000:24.748737pt and 14.142136pt) -- +(45.000000:24.748737pt and 14.142136pt) -- +(135.000000:24.748737pt and 14.142136pt) -- +(225.000000:24.748737pt and 14.142136pt) -- cycle;
\clip (43.500000, -0.000000) +(-45.000000:24.748737pt and 14.142136pt) -- +(45.000000:24.748737pt and 14.142136pt) -- +(135.000000:24.748737pt and 14.142136pt) -- +(225.000000:24.748737pt and 14.142136pt) -- cycle;
\draw (43.500000, -0.000000) node {$DP(\phi)$};
\end{scope}
% Done with gates; drawing ending labels
% Done with ending labels; drawing cut lines and comments
% Done with comments
\end{tikzpicture}

%% file: pics/dp_qudit.tex
\begin{tikzpicture}[scale=1.000000,x=1pt,y=1pt]
\filldraw[color=white] (0.000000, -10.000000) rectangle (79.000000, 90.000000);
% Drawing wires
% Line 1: style=thick a W {a_{+2}} breadth=20
\draw[color=black,thick] (0.000000,80.000000) -- (79.000000,80.000000);
\draw[color=black] (0.000000,80.000000) node[left] {${a_{+2}}$};
% Line 2: style=thick b W {a_{+1}} breadth=20
\draw[color=black,thick] (0.000000,60.000000) -- (79.000000,60.000000);
\draw[color=black] (0.000000,60.000000) node[left] {${a_{+1}}$};
% Line 3: style=thick c W {a_{ 0}} breadth=20
\draw[color=black,thick] (0.000000,40.000000) -- (79.000000,40.000000);
\draw[color=black] (0.000000,40.000000) node[left] {${a_{ 0}}$};
% Line 4: style=thick d W {a_{-1}} breadth=20
\draw[color=black,thick] (0.000000,20.000000) -- (79.000000,20.000000);
\draw[color=black] (0.000000,20.000000) node[left] {${a_{-1}}$};
% Line 5: style=thick e W {a_{-2}} breadth=20
\draw[color=black,thick] (0.000000,0.000000) -- (79.000000,0.000000);
\draw[color=black] (0.000000,0.000000) node[left] {${a_{-2}}$};
% Done with wires; drawing gates
% Line 7: a b c d e / width=10
\draw (6.000000, 74.000000) -- (16.000000, 86.000000);
\draw (6.000000, 54.000000) -- (16.000000, 66.000000);
\draw (6.000000, 34.000000) -- (16.000000, 46.000000);
\draw (6.000000, 14.000000) -- (16.000000, 26.000000);
\draw (6.000000, -6.000000) -- (16.000000, 6.000000);
% Line 9: a G {$P(+2\phi)$} height=20 width=45 fill=red!70!white!50!yellow!25!
\begin{scope}
\draw[fill=red!70!white!50!yellow!25!] (50.500000, 80.000000) +(-45.000000:31.819805pt and 14.142136pt) -- +(45.000000:31.819805pt and 14.142136pt) -- +(135.000000:31.819805pt and 14.142136pt) -- +(225.000000:31.819805pt and 14.142136pt) -- cycle;
\clip (50.500000, 80.000000) +(-45.000000:31.819805pt and 14.142136pt) -- +(45.000000:31.819805pt and 14.142136pt) -- +(135.000000:31.819805pt and 14.142136pt) -- +(225.000000:31.819805pt and 14.142136pt) -- cycle;
\draw (50.500000, 80.000000) node {{$P(+2\phi)$}};
\end{scope}
% Line 10: b G {$P(+1\phi)$} height=20 width=45 fill=red!70!white!50!yellow!25!
\begin{scope}
\draw[fill=red!70!white!50!yellow!25!] (50.500000, 60.000000) +(-45.000000:31.819805pt and 14.142136pt) -- +(45.000000:31.819805pt and 14.142136pt) -- +(135.000000:31.819805pt and 14.142136pt) -- +(225.000000:31.819805pt and 14.142136pt) -- cycle;
\clip (50.500000, 60.000000) +(-45.000000:31.819805pt and 14.142136pt) -- +(45.000000:31.819805pt and 14.142136pt) -- +(135.000000:31.819805pt and 14.142136pt) -- +(225.000000:31.819805pt and 14.142136pt) -- cycle;
\draw (50.500000, 60.000000) node {{$P(+1\phi)$}};
\end{scope}
% Line 11: d G {$P(-1\phi)$} height=20 width=45 fill=red!70!white!50!yellow!25!
\begin{scope}
\draw[fill=red!70!white!50!yellow!25!] (50.500000, 20.000000) +(-45.000000:31.819805pt and 14.142136pt) -- +(45.000000:31.819805pt and 14.142136pt) -- +(135.000000:31.819805pt and 14.142136pt) -- +(225.000000:31.819805pt and 14.142136pt) -- cycle;
\clip (50.500000, 20.000000) +(-45.000000:31.819805pt and 14.142136pt) -- +(45.000000:31.819805pt and 14.142136pt) -- +(135.000000:31.819805pt and 14.142136pt) -- +(225.000000:31.819805pt and 14.142136pt) -- cycle;
\draw (50.500000, 20.000000) node {{$P(-1\phi)$}};
\end{scope}
% Line 12: e G {$P(-2\phi)$} height=20 width=45 fill=red!70!white!50!yellow!25!
\begin{scope}
\draw[fill=red!70!white!50!yellow!25!] (50.500000, -0.000000) +(-45.000000:31.819805pt and 14.142136pt) -- +(45.000000:31.819805pt and 14.142136pt) -- +(135.000000:31.819805pt and 14.142136pt) -- +(225.000000:31.819805pt and 14.142136pt) -- cycle;
\clip (50.500000, -0.000000) +(-45.000000:31.819805pt and 14.142136pt) -- +(45.000000:31.819805pt and 14.142136pt) -- +(135.000000:31.819805pt and 14.142136pt) -- +(225.000000:31.819805pt and 14.142136pt) -- cycle;
\draw (50.500000, -0.000000) node {{$P(-2\phi)$}};
\end{scope}
% Done with gates; drawing ending labels
% Done with ending labels; drawing cut lines and comments
% Done with comments
\end{tikzpicture}

%% file: pics/swap_qudit.qpic.tex
\begin{tikzpicture}[scale=1.000000,x=1pt,y=1pt]
\filldraw[color=white] (0.000000, -7.500000) rectangle (38.000000, 82.500000);
% Drawing wires
% Line 1: color=white x W
\draw[color=white] (0.000000,75.000000) -- (38.000000,75.000000);
% Line 2: style=thick a W {m_{+1}}
\draw[color=black,thick] (0.000000,60.000000) -- (38.000000,60.000000);
\draw[color=black] (0.000000,60.000000) node[left] {$m$};
% Line 3: color=white y W
\draw[color=white] (0.000000,45.000000) -- (38.000000,45.000000);
% Line 4: color=white x1 W
\draw[color=white] (0.000000,30.000000) -- (38.000000,30.000000);
% Line 5: style=thick b W {m_{-1}}
\draw[color=black,thick] (0.000000,15.000000) -- (38.000000,15.000000);
\draw[color=black] (0.000000,15.000000) node[left] {${n}$};
% Line 6: color=white y1 W
\draw[color=white] (0.000000,0.000000) -- (38.000000,0.000000);
% Done with wires; drawing gates
% Line 8: a /
\draw (6.000000, 54.000000) -- (14.000000, 66.000000);
% Line 9: b /
\draw (6.000000, 9.000000) -- (14.000000, 21.000000);
% Line 11: a b SWAP
\draw (29.000000,60.000000) -- (29.000000,15.000000);
\begin{scope}
\draw (26.878680, 57.878680) -- (31.121320, 62.121320);
\draw (26.878680, 62.121320) -- (31.121320, 57.878680);
\end{scope}
\begin{scope}
\draw (26.878680, 12.878680) -- (31.121320, 17.121320);
\draw (26.878680, 17.121320) -- (31.121320, 12.878680);
\end{scope}
% Done with gates; drawing ending labels
% Done with ending labels; drawing cut lines and comments
% Done with comments
\end{tikzpicture}

%% file: pics/swap_qubit.qpic.tex
%! \usetikzlibrary{decorations.pathreplacing,decorations.pathmorphing}
\begin{tikzpicture}[scale=1.000000,x=1pt,y=1pt]
\filldraw[color=white] (0.000000, -7.500000) rectangle (30.000000, 82.500000);
% Drawing wires
% Line 1: a b c W {m_{+1}}<
\draw[color=black] (0.000000,75.000000) -- (30.000000,75.000000);
%   Deferring wire label at (0.000000,75.000000)
% Line 1: a b c W {m_{+1}}<
\draw[color=black] (0.000000,60.000000) -- (30.000000,60.000000);
%   Deferring wire label at (0.000000,60.000000)
% Line 1: a b c W {m_{+1}}<
\draw[color=black] (0.000000,45.000000) -- (30.000000,45.000000);
\filldraw[color=white,fill=white] (0.000000,41.250000) rectangle (-4.000000,78.750000);
\draw[decorate,decoration={brace,amplitude = 4.000000pt},very thick] (0.000000,41.250000) -- (0.000000,78.750000);
\draw[color=black] (-4.000000,60.000000) node[left] {$m$};
% Line 2: d e f W {m_{-1}}<
\draw[color=black] (0.000000,30.000000) -- (30.000000,30.000000);
%   Deferring wire label at (0.000000,30.000000)
% Line 2: d e f W {m_{-1}}<
\draw[color=black] (0.000000,15.000000) -- (30.000000,15.000000);
%   Deferring wire label at (0.000000,15.000000)
% Line 2: d e f W {m_{-1}}<
\draw[color=black] (0.000000,0.000000) -- (30.000000,0.000000);
\filldraw[color=white,fill=white] (0.000000,-3.750000) rectangle (-4.000000,33.750000);
\draw[decorate,decoration={brace,amplitude = 4.000000pt},very thick] (0.000000,-3.750000) -- (0.000000,33.750000);
\draw[color=black] (-4.000000,15.000000) node[left] {$n$};
% Done with wires; drawing gates
% Line 4: a d SWAP
\draw (9.000000,75.000000) -- (9.000000,30.000000);
\begin{scope}
\draw (6.878680, 72.878680) -- (11.121320, 77.121320);
\draw (6.878680, 77.121320) -- (11.121320, 72.878680);
\end{scope}
\begin{scope}
\draw (6.878680, 27.878680) -- (11.121320, 32.121320);
\draw (6.878680, 32.121320) -- (11.121320, 27.878680);
\end{scope}
% Line 5: b e SWAP
\draw (15.000000,60.000000) -- (15.000000,15.000000);
\begin{scope}
\draw (12.878680, 57.878680) -- (17.121320, 62.121320);
\draw (12.878680, 62.121320) -- (17.121320, 57.878680);
\end{scope}
\begin{scope}
\draw (12.878680, 12.878680) -- (17.121320, 17.121320);
\draw (12.878680, 17.121320) -- (17.121320, 12.878680);
\end{scope}
% Line 6: c f SWAP
\draw (21.000000,45.000000) -- (21.000000,0.000000);
\begin{scope}
\draw (18.878680, 42.878680) -- (23.121320, 47.121320);
\draw (18.878680, 47.121320) -- (23.121320, 42.878680);
\end{scope}
\begin{scope}
\draw (18.878680, -2.121320) -- (23.121320, 2.121320);
\draw (18.878680, 2.121320) -- (23.121320, -2.121320);
\end{scope}
% Done with gates; drawing ending labels
% Done with ending labels; drawing cut lines and comments
% Done with comments
\end{tikzpicture}

%% file: pics/mirror_abstract.tex
\begin{tikzpicture}[scale=1.000000,x=1pt,y=1pt]
\filldraw[color=white] (0.000000, -7.500000) rectangle (52.000000, 7.500000);
% Drawing wires
% Line 1: style=thick a W a
\draw[color=black,thick] (0.000000,0.000000) -- (52.000000,0.000000);
\draw[color=black] (0.000000,0.000000) node[left] {$a$};
% Done with wires; drawing gates
% Line 3: a /
\draw (6.000000, -6.000000) -- (14.000000, 6.000000);
% Line 5: a G $M$ width=20 height=20 fill=red!70!white!50!yellow!25!
\begin{scope}
\draw[fill=red!70!white!50!yellow!25!] (36.000000, -0.000000) +(-45.000000:14.142136pt and 14.142136pt) -- +(45.000000:14.142136pt and 14.142136pt) -- +(135.000000:14.142136pt and 14.142136pt) -- +(225.000000:14.142136pt and 14.142136pt) -- cycle;
\clip (36.000000, -0.000000) +(-45.000000:14.142136pt and 14.142136pt) -- +(45.000000:14.142136pt and 14.142136pt) -- +(135.000000:14.142136pt and 14.142136pt) -- +(225.000000:14.142136pt and 14.142136pt) -- cycle;
\draw (36.000000, -0.000000) node {$M$};
\end{scope}
% Done with gates; drawing ending labels
% Done with ending labels; drawing cut lines and comments
% Done with comments
\end{tikzpicture}

%% file: pics/mirror.qpic.tex
\begin{tikzpicture}[scale=1.000000,x=1pt,y=1pt]
\filldraw[color=white] (0.000000, -7.500000) rectangle (46.000000, 67.500000);
% Drawing wires
% Line 1: style=thick a W {m_{+2}}
\draw[color=black,thick] (0.000000,60.000000) -- (46.000000,60.000000);
\draw[color=black] (0.000000,60.000000) node[left] {${a_{+2}}$};
% Line 2: style=thick b W {m_{+1}}
\draw[color=black,thick] (0.000000,45.000000) -- (46.000000,45.000000);
\draw[color=black] (0.000000,45.000000) node[left] {${a_{+1}}$};
% Line 3: style=thick c W {m_{ 0}}
\draw[color=black,thick] (0.000000,30.000000) -- (46.000000,30.000000);
\draw[color=black] (0.000000,30.000000) node[left] {${a_{ 0}}$};
% Line 4: style=thick d W {m_{-1}}
\draw[color=black,thick] (0.000000,15.000000) -- (46.000000,15.000000);
\draw[color=black] (0.000000,15.000000) node[left] {${a_{-1}}$};
% Line 5: style=thick e W {m_{-2}}
\draw[color=black,thick] (0.000000,0.000000) -- (46.000000,0.000000);
\draw[color=black] (0.000000,0.000000) node[left] {${a_{-2}}$};
% Done with wires; drawing gates
% Line 7: a b c d e / width=10
\draw (6.000000, 54.000000) -- (16.000000, 66.000000);
\draw (6.000000, 39.000000) -- (16.000000, 51.000000);
\draw (6.000000, 24.000000) -- (16.000000, 36.000000);
\draw (6.000000, 9.000000) -- (16.000000, 21.000000);
\draw (6.000000, -6.000000) -- (16.000000, 6.000000);
% Line 9: a e SWAP
\draw (31.000000,60.000000) -- (31.000000,0.000000);
\begin{scope}
\draw (28.878680, 57.878680) -- (33.121320, 62.121320);
\draw (28.878680, 62.121320) -- (33.121320, 57.878680);
\end{scope}
\begin{scope}
\draw (28.878680, -2.121320) -- (33.121320, 2.121320);
\draw (28.878680, 2.121320) -- (33.121320, -2.121320);
\end{scope}
% Line 10: b d SWAP
\draw (37.000000,45.000000) -- (37.000000,15.000000);
\begin{scope}
\draw (34.878680, 42.878680) -- (39.121320, 47.121320);
\draw (34.878680, 47.121320) -- (39.121320, 42.878680);
\end{scope}
\begin{scope}
\draw (34.878680, 12.878680) -- (39.121320, 17.121320);
\draw (34.878680, 17.121320) -- (39.121320, 12.878680);
\end{scope}
% Done with gates; drawing ending labels
% Done with ending labels; drawing cut lines and comments
% Done with comments
\end{tikzpicture}

%% file: pics/hologram_abstract.tex
\begin{tikzpicture}[scale=1.000000,x=1pt,y=1pt]
\filldraw[color=white] (0.000000, -7.500000) rectangle (52.000000, 7.500000);
% Drawing wires
% Line 1: style=thick a W a
\draw[color=black,thick] (0.000000,0.000000) -- (52.000000,0.000000);
\draw[color=black] (0.000000,0.000000) node[left] {$a$};
% Done with wires; drawing gates
% Line 3: a /
\draw (6.000000, -6.000000) -- (14.000000, 6.000000);
% Line 5: a G $G$ width=20 height=20 fill=red!70!white!50!yellow!25!
\begin{scope}
\draw[fill=red!70!white!50!yellow!25!] (36.000000, -0.000000) +(-45.000000:14.142136pt and 14.142136pt) -- +(45.000000:14.142136pt and 14.142136pt) -- +(135.000000:14.142136pt and 14.142136pt) -- +(225.000000:14.142136pt and 14.142136pt) -- cycle;
\clip (36.000000, -0.000000) +(-45.000000:14.142136pt and 14.142136pt) -- +(45.000000:14.142136pt and 14.142136pt) -- +(135.000000:14.142136pt and 14.142136pt) -- +(225.000000:14.142136pt and 14.142136pt) -- cycle;
\draw (36.000000, -0.000000) node {$G$};
\end{scope}
% Done with gates; drawing ending labels
% Done with ending labels; drawing cut lines and comments
% Done with comments
\end{tikzpicture}

%% file: pics/hologram.qpic.tex
\begin{tikzpicture}[scale=1.000000,x=1pt,y=1pt]
\filldraw[color=white] (0.000000, -7.500000) rectangle (72.000000, 67.500000);
% Drawing wires
% Line 1: style=thick a W {m_{+2}}
\draw[color=black,thick] (0.000000,60.000000) -- (72.000000,60.000000);
\draw[color=black] (0.000000,60.000000) node[left] {${a_{+2}}$};
% Line 2: style=thick b W {m_{+1}}
\draw[color=black,thick] (0.000000,45.000000) -- (72.000000,45.000000);
\draw[color=black] (0.000000,45.000000) node[left] {${a_{+1}}$};
% Line 3: style=thick c W {m_{ 0}}
\draw[color=black,thick] (0.000000,30.000000) -- (72.000000,30.000000);
\draw[color=black] (0.000000,30.000000) node[left] {${a_{ 0}}$};
% Line 4: style=thick d W {m_{-1}}
\draw[color=black,thick] (0.000000,15.000000) -- (72.000000,15.000000);
\draw[color=black] (0.000000,15.000000) node[left] {${a_{-1}}$};
% Line 5: style=thick e W {m_{-2}}
\draw[color=black,thick] (0.000000,0.000000) -- (72.000000,0.000000);
\draw[color=black] (0.000000,0.000000) node[left] {${a_{-2}}$};
% Done with wires; drawing gates
% Line 7: a b SWAP
\draw (9.000000,60.000000) -- (9.000000,45.000000);
\begin{scope}
\draw (6.878680, 57.878680) -- (11.121320, 62.121320);
\draw (6.878680, 62.121320) -- (11.121320, 57.878680);
\end{scope}
\begin{scope}
\draw (6.878680, 42.878680) -- (11.121320, 47.121320);
\draw (6.878680, 47.121320) -- (11.121320, 42.878680);
\end{scope}
% Line 8: b c SWAP
\draw (27.000000,45.000000) -- (27.000000,30.000000);
\begin{scope}
\draw (24.878680, 42.878680) -- (29.121320, 47.121320);
\draw (24.878680, 47.121320) -- (29.121320, 42.878680);
\end{scope}
\begin{scope}
\draw (24.878680, 27.878680) -- (29.121320, 32.121320);
\draw (24.878680, 32.121320) -- (29.121320, 27.878680);
\end{scope}
% Line 9: c d SWAP
\draw (45.000000,30.000000) -- (45.000000,15.000000);
\begin{scope}
\draw (42.878680, 27.878680) -- (47.121320, 32.121320);
\draw (42.878680, 32.121320) -- (47.121320, 27.878680);
\end{scope}
\begin{scope}
\draw (42.878680, 12.878680) -- (47.121320, 17.121320);
\draw (42.878680, 17.121320) -- (47.121320, 12.878680);
\end{scope}
% Line 10: d e SWAP
\draw (63.000000,15.000000) -- (63.000000,0.000000);
\begin{scope}
\draw (60.878680, 12.878680) -- (65.121320, 17.121320);
\draw (60.878680, 17.121320) -- (65.121320, 12.878680);
\end{scope}
\begin{scope}
\draw (60.878680, -2.121320) -- (65.121320, 2.121320);
\draw (60.878680, 2.121320) -- (65.121320, -2.121320);
\end{scope}
% Done with gates; drawing ending labels
% Done with ending labels; drawing cut lines and comments
% Done with comments
\end{tikzpicture}